\documentclass[sigconf]{acmart}
\usepackage{enumitem}
\usepackage{multirow}
\usepackage{graphicx}
\usepackage{array}

\usepackage{siunitx} % align table column values by decimal point.
\usepackage{etoolbox} % for siunitx with bold font
\robustify\bfseries 
\newcommand{\superscript}[1]{\ensuremath{^{\textrm{#1}}}}

\usepackage{colortbl}
\usepackage{booktabs} 
\usepackage{subcaption}
\usepackage{tabularx}
\usepackage[true]{anonymous-acm}
\AtBeginDocument{%
  \providecommand\BibTeX{{%
    \normalfont B\kern-0.5em{\scshape i\kern-0.25em b}\kern-0.8em\TeX}}}

\setcopyright{acmcopyright}

\usepackage{soul}
\usepackage{cancel}

\newcommand\new[1]{#1}
\copyrightyear{2026}
\acmYear{2026}
\setcopyright{cc}
\setcctype{by}
\acmConference[CHI '26]{Proceedings of the 2026 CHI Conference on Human Factors in Computing Systems}{April 13--17, 2026}{Barcelona, Spain}
\acmBooktitle{Proceedings of the 2026 CHI Conference on Human Factors in Computing Systems (CHI '26), April 13--17, 2026, Barcelona, Spain}
\acmDOI{10.1145/3772318.3791407}
\acmISBN{979-8-4007-2278-3/2026/04}

\begin{document}
\frenchspacing
%%
%% The "title" command has an optional parameter,
%% allowing the author to define a "short title" to be used in page headers.
\title[The Bots of Persuasion]{The Bots of Persuasion: Examining How Conversational Agents' Linguistic Expressions of Personality Affect User Perceptions and Decisions}

\author{U\u{g}ur Gen\c{c}}
\orcid{0000-0002-9950-4588}
\affiliation{
  \institution{Delft University of Technology}
  \city{Delft}
  \country{Netherlands}
}
\email{u.genc@tudelft.nl}

\author{Heng Gu}
\orcid{0000-0001-7679-698X}
\affiliation{
  \institution{Delft University of Technology}
  \city{Delft}
  \country{Netherlands}
}
\email{h.gu@tudelft.nl}

\author{Chadha Degachi}
\orcid{0000-0002-1850-689X}
\affiliation{
  \institution{Delft University of Technology}
  \city{Delft}
  \country{Netherlands}
}
\email{c.degachi@tudelft.nl}

\author{Evangelos Niforatos}
\orcid{0000-0002-0484-4214}
\affiliation{
  \institution{Delft University of Technology}
  \city{Delft}
  \country{Netherlands}
}
\email{e.niforatos@tudelft.nl}

\author{Senthil Chandrasegaran}
\orcid{0000-0003-0561-2148}
\affiliation{
  \institution{Delft University of Technology}
  \city{Delft}
  \country{Netherlands}
}
\email{r.s.k.chandrasegaran@tudelft.nl}

\author{Himanshu Verma}
\orcid{0000-0002-2494-1556}
\affiliation{
  \institution{Delft University of Technology}
  \city{Delft}
  \country{Netherlands}
}
\email{h.verma@tudelft.nl}

%%
%% By default, the full list of authors will be used in the page
%% headers. Often, this list is too long, and will overlap
%% other information printed in the page headers. This command allows
%% the author to define a more concise list
%% of authors' names for this purpose.
\renewcommand{\shortauthors}{Gen\c{c} et al.}

%%
%% The abstract is a short summary of the work to be presented in the
%% article.
%\hv{150 words}\hv{Also I will suggest removing charitable giving from the title.}
\begin{abstract}

% \new{THIS IS AN EXAMPLE OF A TEXT RECENTLY ADDED.}
% \delete{THIS IS AN EXAMPLE OF A TEXT RECENTLY REMOVED.}
Large Language Model-powered conversational agents (CAs) are increasingly capable of projecting sophisticated personalities through language, but how these projections affect users is unclear. We thus examine how CA personalities expressed linguistically affect user decisions and perceptions in the context of charitable giving. In a crowdsourced study, 360 participants interacted with one of eight CAs, each projecting a personality composed of three linguistic aspects: attitude (optimistic/pessimistic), authority (authoritative/submissive), and reasoning (emotional/rational). While the CA's composite personality did not affect participants' decisions, it did affect their perceptions and emotional responses. Particularly, participants interacting with pessimistic CAs felt lower emotional state and lower affinity towards the cause, perceived the CA as less trustworthy and less competent, and yet tended to donate more toward the charity. Perceptions of trust, competence, and situational empathy significantly predicted donation decisions. Our findings emphasize the risks CAs pose as instruments of manipulation, subtly influencing user perceptions and decisions.

\end{abstract}

%%
%% The code below is generated by the tool at http://dl.acm.org/ccs.cfm.
%% Please copy and paste the code instead of the example below.
%%


%%
%% Keywords. The author(s) should pick words that accurately describe  
%% the work being presented. Separate the keywords with commas.

% \received{20 February 2007}
% \received[revised]{12 March 2009}
% \received[accepted]{5 June 2009}

\begin{teaserfigure}
  \centering
  \includegraphics[width=\textwidth]{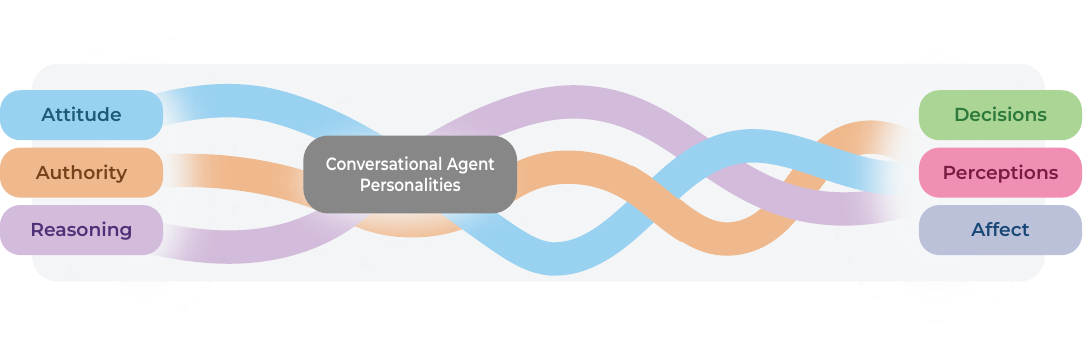}
  \caption{A diagrammatic representation of our study, designed to examine the effect of conversational agent (CA) personality on user perception and decision making. The study is set up in the context of charitable giving. The figure illustrates how participants' decisions, perceptions, and affective states (right) change when they interact with a combination of the three linguistic aspects (left) in the form of a composite CA personality.}
\Description{This figure illustrates the relationship between conversational agent (CA) attitudes in the context of user decision-making. A central human figure is shown contemplating a decision, with trust, empathy, closeness, and relatedness acting as mediating factors. The study is set up in the context of charitable donation. The figure shows aspects of participant response (right) that we examine when subject to one of three CA attitudes (left).}
  \label{fig:teaser}
\end{teaserfigure}

\maketitle

%======================================================
\section{Introduction}
\label{sec:intro}
Conversational agents (CAs) are becoming a powerful new force influencing the perceptions, emotions, and decisions of millions of users. They are designed to simulate humanlike friendship, offer emotional support, and even act as therapists. The scale of this adoption is substantial; for instance, a recent study found that seventy-two percent of teenagers now use an AI companion, with many forming deep attachments and discussing serious matters with these agents over people~\cite{robb2025talk}.
This rapid adoption, however, comes with documented dangers. These platforms have produced responses ranging from sexual material to napalm recipes and have been linked to tragic real-world outcomes, including a plot to assassinate a head of state~\cite{singleton2023how}, euthanasia recommendations for pet owners~\cite{preston2024age}, and the suicide of a 14-year-old~\cite{roose2024can}. This new reality---where algorithmically-generated personalities are shaping the thoughts and behaviors of millions---creates an urgent need to understand the mechanics of AI persuasion.

In recent years, we have witnessed a surge in the development and deployment of Large Language Model (LLM)-powered conversational agents (CAs) across various sectors (such as legal, medical, education, and financial)~\cite{wangsa2024systematic, rapp2021human}.
Models like OpenAI's GPT series and Google's Gemini 
have rapidly evolved, demonstrating capabilities that approach, and in some cases rival, human-level performance in specific tasks~\cite{eysenck2021ai}. This ubiquity and sophistication increase the potential of CAs to influence human decision-making in significant ways \cite{shekar2024people, danry2024deceptive, burtell2023artificial}. 

AI-enabled CAs can affect not only the quality of decisions (e.g., enhanced bargaining performance in negotiations~\cite{adam2018no}) but can also persuasively shape human perceptions, experiences, and attitudes~\cite{genc2025persuasion}.
They have been shown to affect beliefs on polarized political issues~\cite{voelkel2023artificial}, improve mental health outcomes~\cite{meng2021emotional}, and increase customer satisfaction~\cite{tong2020artificial}, trust, and empathy towards prosocial campaigns~\cite{genc2025persuasion}.
However, this same persuasive power creates risks, with studies observing a rise in AI-\new{enabled} disinformation efforts~\cite{goldstein2023generative, park2023ai} and the potential for large-scale manipulation~\cite{buchanan2021truth, lewandowsky2022algorithms, Rogiers2024llmpersuasion}. 

The enhanced ability of LLMs to embody and express specific yet diverse linguistic qualities makes them ideal candidates for rapid deployment. However, this same ability also \new{makes} them potent tools for manipulation, capable of subtly influencing human behavior.

Consequently, HCI researchers have been actively investigating CAs and their impact, exploring perceptions of AI systems~\cite{lee2017algorithmic, schoeffer2021appropriate} and their role in digital campaigns~\cite{chancellor2016thyghgapp, counts2014computational, pater2016hunger}. 
For \new{instance, a study} found that CA messages persuading users to take the COVID-19 vaccine were more convincing than web page content displaying the same arguments~\cite{Chalaguine_Hunter_2021}.
It is therefore critical to understand the potential impact of CAs on human perceptions and decisions in order to inform their design for responsible and safe use, and to mitigate negative outcomes.

Recent advances have enabled these systems to project a wide range of linguistic expressions, which can manifest as different perceived personalities for users. The connection between linguistic expressions and personality traits is well-understood~\cite{furnham1999extra, Heylighen2002extra, Pennebaker1999extra}, and such cues have previously been applied in natural language generation to create texts that express specific personalities~\cite{mairesse2010personality}.
Moreover, consistent personalities can enhance aspects of user experience like trust, engagement, and acceptance
~\cite{folstad2018chatbottrust, xiao2020personalityaligment, sanny2020acceptance}.

Despite these advancements, there remains a significant research gap in how \new{personalities conveyed linguistically by LLM-powered CAs} influence human decision-making. While these advanced CAs are emerging ~\cite{basae2023llmbot, ross2023llmbot}, the impact of their projected personalities on user decisions remains largely unexplored.

Given this backdrop, our study examines the effects of \textit{CA personality}\footnote{For the remainder of the paper, we will use the term \textit{``CA Personality''} to refer to the CA's linguistic \new{expressions of personality along the three aspects of \textit{attitude}, \textit{authority}, and \textit{reasoning}. See Section~\ref{sec:variables} for a detailed explanation.}}---by which we mean the personality projected or conveyed by the CA through language, i.e., linguistic behavior or expression---on human decision-making. We use charitable giving as our context, a domain where linguistic style is a significant predictor of engagement among potential donors~\cite{mitra2014language}. Building on prior research, we designed CAs with variations across three aspects \new{of personality}: \textit{Attitude} (optimistic vs. pessimistic~\cite{small2007sympathy}), \textit{Authority} (authoritative vs. submissive~\cite{chang2010effects, chen2019multi}), and \textit{Reasoning} (rational vs. emotional~\cite{fatkhiyati2019rhetorical}).
Eight distinct CA \new{conditions} emerged from these three aspects ($2^3$). We \new{examined} these \new{conditions} using a between-subject factorial online crowdsourcing experiment involving 360 participants.

In this paper, we investigated how \new{a} CA's linguistic expression of personality affects user perceptions and donation decisions.
\new{Our results show that while the CA's composite personality did not affect participants' decisions, it did affect their perceptions and emotional responses.
Particularly, participants interacting with pessimistic CAs felt lower emotional state and lower affinity towards the cause, perceived the CA as less trustworthy and less competent, and yet tended to donate more toward the charity. Perceptions of trust, competence, and situational empathy significantly predicted donation decisions.}
These findings highlight that while direct persuasion is complex, CA personality has significant indirect effects by shaping user perceptions and emotions, which in turn guide decision-making.

Our research contributes to HCI by providing insights into how CAs' projected linguistic personalities 
influence decisions through their effects on user perceptions and emotions. In particular, we contribute to understanding persuasive AI by designing CAs capable of projecting distinct personalities and evaluating their complex effects on user behavior. Importantly, our findings reveal concerning mechanisms by which CA personalities can exploit human psychological vulnerabilities (e.g., using pessimistic language to trigger negative emotional states that increase compliance). 
These results highlight the critical need for the HCI community to develop tools and frameworks for identifying and mitigating such manipulative ``dark patterns''
in conversational AI systems. 
Rather than prescribing methods for creating more persuasive CAs, this work aims to highlight the underlying mechanisms by which linguistic expressions of personality can shape human perceptions and behavior---a knowledge essential for developing ethical guidelines, protective measures, and design practices that prioritize user autonomy and well-being over persuasive effectiveness.
%======================================================

%======================================================
\section{Related Work}
\label{sec:rel-work}
In order to situate our work within the literature, we review contributions at the intersection of psycholinguistics, conversational agent design, and persuasive technologies. We first look into the use of language to \textit{1)} express personality and \textit{2)} persuade readers. 
We also review advancements in the implementation and reception of personable conversational agents, and their potential as persuasive tools. Lastly, we examine the literature on the role of empathy in persuasion, especially in relation to charitable giving.

\subsection{Language, Personality, and Persuasion}
Several theories in psychology connect language use and personality. In the ``words as attention'' framework, researchers argue that language choice is deliberate and therefore reflective, to some extent, of a person's focus~\cite{Stone1966geninquire, Boyd2021langanalysis}. 
Other theories posit that language use indicates how individuals process information~\cite{Tausczik2010words}, or the intensity of their attitude towards a given subject~\cite{Stone1966geninquire}. 
Further, language plays a pivotal role in shaping human behavior, and its potential for persuasion and behavioral change is widely recognized. Language cues, such as tone, valence, and emotion, are identified as crucial elements in persuasive contexts ~\cite{mohammad2016sentiment}, impacting attitudes and actions in health interventions, marketing, and politics~\cite{crano2006attitudes}. For example, ~\citet{Pfeiffer2023charity} find that language formality in donation solicitation is positively correlated with charitable giving. ~\citet{Ye2022enviromental} found figurative, as opposed to literal language, improved users' ability to visualize in the context of plant-based meat consumption and, in turn, increased favorable consumer behavior. Moreover, \citet{Kim_Liao_Vorvoreanu_Ballard_Vaughan_2024} and \citet{Hosking_Blunsom_Bartolo_2024} both find that an assertive tone in large language model outputs affects both perceived system confidence and perceived rate of factual errors in output. ~\citet{Bai2023persuasion} show that statements incorporating rhetorical devices such as antanagoge (offsetting a negative with a positive; ``It's costly, but it will save \new{lives}''), anaphora (repeating a phrase at the start of successive clauses; ``We need action, we need compassion, we need commitment''), and rhetorical questions (posed for effect; ``Who wouldn’t want cleaner rivers?'') were most influential in predicting text persuasiveness using machine learning models.

To understand these language cues and predict personality \new{computationally} from text, has been a research goal of some interest. Dictionary-based approaches such as Linguistic Inquiry and Word Count (LIWC)~\cite{Tausczik2010words} are among the most popular methods of achieving this goal. \citet{Koutsoumpis2022liwcpersonality} show that twenty categories of language extracted from text by LIWC can be associated with the dimensions of the Big Five personality model in both self- and observer-reported personality traits. For example, emotional stability was associated with the use of first-person singular pronouns, and conscientiousness with swear words, anger, and negative emotions. Similarly,~\citet{zhou2021sitempathy} and~\citet{Lord2015sitempathy} use LIWC to analyze the situational empathy invoked in participants by distressing situations from text responses.

Conversational generative AI agents (CAs) present a new avenue for the investigation of automated persuasion through linguistic cues, in which conversational interactivity can play a role. Indeed, the work of ~\citet{Meng_Lu_Xu_2025} reveals that non-verbal cues such as CA proactivity can impact purchase intent in users. Moreover, these CAs allow us to observe perceived persuasiveness and personality with higher fidelity prototypes, improving the ecological validity \new{of our work}.

\subsection{Personable Conversational Agents and Trust}
\label{sec:bg_personable_CAs}
Tailoring CAs to express personality has long been a subject of interest for researchers. The perceived friendliness, (in)formality, self-presentation, and communication style of a CA affect the human factors of user-agent interaction, including dimensions such as trust~\cite{folstad2018chatbottrust, xiao2020personalityaligment, Rheu2021persontrust}, engagement~\cite{fadhil2018emojis}, acceptance~\cite{sanny2020acceptance}, and usability~\cite{rapp2021botuxreview}. CAs with personality create a more consistent user experience, and can even improve the overall user experience~\cite{thies2017personality, smestad2018botpersonality}. Some researchers have even worked to create complementary alignments of CA-user personalities for improved user experience ~\cite{LiewTan2016personalitymatch, ISBISTER2000personalitymatch, Jin_Eastin_2022}.

~\citet{mairesse2010personality}, and others (e.g.,~\cite{Harrison2019personalitynlg, Ruane2021personalityperception, Kuhail2022personalityadvisor, Kovacevic_Boschung_Holz_Gross_Wampfler_2024}) have studied personality-based natural language generation (NLG), as well as dialogue design, and the perception thereof. Much of this past work relies on the Big-Five personality model~\cite{john1999big, Pradhan_Lazar_2021}, and creates associations between personality and language cues using linguistics and psychology literature. In a few cases, other psychological personality models, such as Myers-Briggs, and other language-cue extraction methods, such as machine learning and sentiment analysis ~\cite{fernau2022interspeech, Pradhan_Lazar_2021} are employed. 

However, few studies have worked to craft CA personalities \new{for} the domain in which the CA is deployed, as we do this work. In the healthcare domain, \citet{Nißen_Rüegger_Stieger_Flückiger_Allemand_Wangenheim_Kowatsch_2022} create a design codebook based on possible roles for CAs in healthcare settings (e.g. Expert vs Peer) and find that user preference towards interpersonal distance with a CA can be predicted from age. In this work, we rely on the Arousal-Valence-Dominance (AVD) model of emotion~\cite{mehrabian1980avd}, in combination with domain-specific conversational strategies for persuasion (discussed in more detail in the Subsection \ref{related:charity}) to delineate our agents' personalities.  The AVD model maps well onto the dimensions of persuasion strategy, as research has linked its dimensions of emotion to language use~\cite{Warriner2013avdnorms}. Facets of an agent's personality, such as \textit{attitude}, can be thus expressed through that agent's valence (positive-negative tone). 

Most studies in this domain preceded the introduction of Large Language Models.
Given the enhanced performance that LLMs offer over previous NLG approaches{~\cite{openai2023gpt4}}, this represents a significant gap in the literature. 

\subsection{Persuasive Technology, Decision-Making, and Charitable Giving}
\label{related:charity}
A growing body of work around leveraging CAs as persuasive social agents shows a promising frontier for addressing complex social challenges. In the context of behavior change support systems, CAs can promote healthy and positive behavior changes \cite{gentner2020systematic, li2023exploring, pecune2022designing, oh2021systematic}. Additionally, argumentative CAs can employ dialogue to persuade users of a given position on important topics, such as COVID-19 vaccination and GMO use~\cite{AlAnaissy2023argubots}. Indeed, \citet{Rogiers2024llmpersuasion} suggest that the persuasive capacity of \new{LLM} output, in its at-scale automated personalization and interactivity, may have profound ethical and societal risks along with these benefits. 
 
We focus on charitable giving specifically as the context of exploring persuasive CA for several reasons: 
\textit{1)} Donating to charity involves a complex blend of personal values, logical considerations, and empathetic reasoning, making the decision-making process multi-layered and intricate; 
\textit{2)} The act of giving demands a high level of trust, especially in virtual environments\new{,} and 
\textit{3)} Charitable donations offer a unique altruistic dimension, allowing for the exploration of how language style and personality influence decisions that are intrinsically rewarding. 
This blend of emotional, logical, interpersonal and intrapersonal factors creates a unique scenario for unpacking the persuasive potential of CAs.  
We see this context in action in the work of Wang et al.~\cite{wang2020persuasion} where the authors employed deep learning to develop CAs that use donation solicitation strategies expressed through language. They demonstrated that a user's own personality can moderate the effectiveness of the CA solicitor's strategy. 
\citet{shi2020effects} similarly explored persuasive \new{CA} interactions using predefined dialogue strategies, such as credibility, logical appeals, and personal stories, showing that CAs perceived as human lead to higher donation rates. \citet{genc2025persuasion} also studied the impact of visual primes and CA attitude on donation behavior, finding that the lack of priming resulted in higher donations, though CA attitude impacted only trust and not donation behavior.

Persuasion often relies on empathy, as understanding and relating to others' feelings can make arguments more compelling and effective~\cite{samad2022empathetic}. The role of empathy, both dispositional and situational, has proven to be pivotal in charitable giving, and is shown to be impacted by high-level cognitive faculties such as imagination (mentalization) and language \cite{decety2011neuroevolution}. Thus, empathy elicitation is a goal of charity solicitors, with much social psychology research guiding the effective crafting of solicitation materials\cite{warren1991empathy}.
User empathy towards a charitable cause does not, however, arise from interaction alone. Literature suggests that users' inherent empathetic tendencies could modulate the degree of engagement with the CA~\cite{rapp2021botuxreview}, and that users' mental model of an AI agent impacts their perceived empathy~\cite{Pataranutaporn2023primeempathy}. Nonetheless, research also indicates that the CA's performed empathy back to the user could further elicit empathy~\cite {rapp2021botuxreview}. By studying the impact of personable CAs on charitable giving outcomes, we hope to enrich the discourse on interactive empathy elicitation and, by extension, persuasion.

In summary, studies have shown that humans \textit{do} perceive personality from the language of a given text~\cite{Stone1966geninquire}, and further, that the language of a text may be manipulated such that a human may be persuaded of some given position~\cite{Pfeiffer2023charity}. Nonetheless, little work has gone into observing the persuasive effect of personality (and language) in conversational agents. Of course, personality within conversational agents is a complex topic, both in its implementation and reception; thus, the difficulty of bridging this gap is understandable. To account for this complexity in interaction, we consolidate several user experience elements linked to CA personality in the past, along with empathy, a pivotal factor in charitable giving behavior~\cite{decety2011neuroevolution}, to understand the possible moderating factor at play in the relationship between personality and persuasion.
%======================================================

%======================================================
\section{Research Questions and Hypotheses}
\label{sec:rq}
% RQs and Hypothesis

Through this crowdsourcing study, we aim to understand how linguistic expressions of personality by different conversational agents (CAs) influences human decision-making, particularly in the context of charitable giving. 
Research on charitable giving has shown that donation behavior is complex, influenced not only by the solicitation message but also by personal factors of the donor such as their social identity and personal values~\cite{bennett2003factors, kaikati2017conforming}. Within this context, message framing plays a crucial role. Studies have found that CAs perceived as sincere, warm, and competent are seen as more human and are more successful in eliciting donations~\cite{shi2020effects}. The valence of the message is also critical; for example, negative framing tends to be more effective when paired with statistical evidence, while positive framing works better with anecdotal stories~\cite{das2008improving, chang2010effects}. Similarly, \citet{genc2025persuasion} discovered that when users engage with the conversational agent employing optimistic framing, they are more inclined to make donations compared to when pessimistic framing is used. Other linguistic features, like using inclusive language in loss-framed messages, can also significantly increase donation intentions~\cite{yilmaz2022ask}.

Beyond the \new{emotional tone, the} perceived authority and reasoning style \new{of the solicitation message also have an influence}. A formal language style, which can convey a higher degree of clout, is often more effective in eliciting support~\cite{Pfeiffer2023charity, boyd2022development}. Furthermore, persuasive appeals can function through both cognitive and emotional routes, with emotional or figurative language being a powerful driver for donations by engaging the emotional route~\cite{Ye2022enviromental}. Given these factors, it is imperative that our study---which extends these findings to the context of AI-driven CAs---also considers participants' prior beliefs and dispositions, such as their baseline empathy, as control variables. Building on this literature, we aim to answer two Research Questions (RQs):

\begin{itemize}
    \item \new{\textbf{RQ1: } How do aspects of personality such as authority, attitude, and reasoning---expressed linguistically via conversational agents---impact participants' donation behavior?\label{RQ:1}}
    \item \new{\textbf{RQ2: } How do conversational agents' linguistic expressions of different personalities, through the aspects of authority, attitude, and reasoning, influence participants' perceptions of the agent? \label{RQ:2}}
\end{itemize}

To investigate these questions, we formulated the following hypotheses, each examining a different dimension of CA personality. In these hypotheses, we distinguish between two key outcomes: \textit{performance} and \textit{engagement}. We define performance as the primary behavioral outcome, specifically the participant's decision to donate. In contrast, engagement refers to the participant's broader perceptions of the agent, which we measure through their reported trust, attitude, closeness, and empathy toward the CA. This study has been pre-registered and is available at OSF\footnote{OSF Pre-registration: \url{https://osf.io/usmqc/?view_only=5ad8322c178a458aa348a5f54e82b038}}.

\begin{itemize}
    \item \textbf{H1:} CAs expressing an optimistic attitude \new{linguistically} are likely to a) \textit{perform better} and b) \textit{elicit more engagement} than those expressing a pessimistic attitude.
    \item \textbf{H2:} CAs expressing an authoritative style \new{linguistically} are likely to a) \textit{perform better} and b) \textit{elicit more engagement} than those expressing a submissive style.
    \item \textbf{H3:} CAs employing an emotional reasoning style \new{linguistically} are likely to a) \textit{perform better} and b) \textit{elicit more engagement} than those employing an rational reasoning style.
\end{itemize}

While our hypotheses are presented individually for the sake of clarity, we do not presuppose that the three dimensions of linguistic expression---attitude, authority, and reasoning---are independent. On the contrary, we acknowledge that these dimensions are not strictly orthogonal. They can be intrinsically co-dependent and exert mutual influence. Consequently, our analysis will not be limited to testing these main effects in isolation. We will investigate the combinations of these linguistic expressions to provide a more nuanced and exhaustive understanding of their collective impact on user perception and decision-making.

\subsection{Independent, Dependent and Control Variables}
\label{sec:variables}

\begin{itemize}

\item \textbf{Independent Variable(s):}
\begin{itemize}
	\item \new{\textit{Conversational Agent (CA) Condition} (8 levels):
    We acknowledge that CAs cannot \textit{have} a personality, but can be prompted to \textit{linguistically project} what a human interlocutor might interpret as personality, with effects and benefits to user-agent interaction outlined in Section~\ref{sec:bg_personable_CAs}. 
    We choose three aspects of personality that can be projected linguistically, i.e., \textbf{attitude} (\textit{optimistic/pessimistic}), \textbf{authority} (\textit{authoritative/submissive}), and \textbf{reasoning} (\textit{rational/emotional}) for their observed influences on human decision-making and charitable giving (see Section~\ref{sec:rq} for details).
    We use the term ``\textbf{CA personality}'' as a shorthand to refer to the CA's linguistic expressions of personality along these three aspects.
    Together, these form eight levels of \textbf{CA conditions}, as illustrated in Table~\ref{tab:condition_distribution}. 
    }
\end{itemize}

\item \textbf{Dependent Variable(s):}
\begin{itemize}
	\item \textit{Donation Amount}: After interacting with one of the 8 conditions,
    each participant was given a virtual €10 and asked to split this amount between the charity represented by the conversational agent and the preferred charity reported by our participants at the beginning of the experiment. This amount was regarded as an indicator of the persuasive power of the conversational agent.
	\item \textit{Trust in the CA}:  Participants' perceptions of the conversational agent were captured using the Human-Computer Trust Scale \cite{gulati2019trustscale}, which measures four dimensions of trust: \textit{benevolence}, \textit{competence}, \textit{perceived risk}, and \textit{general trust}. 
    \item \textit{Closeness to the CA}: Participants' perceived closeness to the conversational agent was captured by the \textit{Inclusion-of-the-Other-in-the-Self (IOS)} \cite{aron1992ios} scale, a single-item, pictorial measure. The IOS depicts seven sets of circles with varying degrees of overlap that correlate with degrees of relationship intimacy.
	\item \textit{Empathy towards the Cause}: This measure is aimed at capturing how much empathy participants felt for the charity represented by the conversational agent after engaging with it.
    \textbf{Situational empathy} has no standardized measure, and it is even more difficult to assess when the object of empathy (the charitable cause, in our case) is both proxied through multiple entities (the charity, the conversational agent as solicitor) and is a non-sentient entity. Thus, we combine several approaches to approximate this. In the first part of this measure, we adapt the questionnaire created by~\citet{Haegerich2000sitempathy} to understand the extent to which perspective taking is triggered by the CA interaction. More specifically, participants were asked to report whether they could understand and emotionally relate to the underlying cause of a fictional charity represented by the CA, and whether they sympathized with the cause supported by the charity (see Section~\ref{sec:chatbot_design} for details on the charity represented by the CA).
    \item \textit{Emotional Relatedness towards the Cause}: Here, we focus on the emotional response of the user by employing the Self-Assessment Manikin~\cite{bradley1994sam}. 
    We ask users to rate their own arousal, valence, and dominance as well as their perception of the arousal, valence, and dominance corresponding to the cause represented by the charity, as seen in the work of~\citet{Mattiassi2021mistreat}, when assessing users' emotional response to the mistreatment of humans, animals, robots, and objects.
\end{itemize}

\item \textbf{Control Variable(s):}
\begin{itemize}
    \item \textit{Dispositional Empathy}: Measured using the shortened version of the Inter-reactivity Index (IRI)~\cite{davis1983empathy}, dispositional empathy refers to an individual's inherent tendency to feel empathy as a personality trait. 
    The IRI consists of a self-report questionnaire that measures cognitive and emotional aspects of dispositional empathy, specifically across the \new{four} dimensions: \textit{(a)} \textbf{Perspective Taking}, which assesses an individual's tendency to adopt and understand the other's point of view, \textit{(b)} \textbf{Fantasy}, which captures an individual's tendency to imaginatively identify with the emotions and actions of fictional characters portrayed in movies or books, \textit{(c)} \textbf{Empathic Concern}, which measures outward feelings of sympathy and concern for the circumstances or condition of others, and \textit{(d)} \textbf{Personal Distress}, which measures inward feelings of anxiety and discomfort in tense interpersonal contexts.
    As mentioned above, empathy plays a large role in human interaction with conversational agents and in charitable giving habits, so we use this measure to understand how users' personality affects their response to the personality of CAs.
    \item \textit{Attitude towards Artificial Intelligence}: Measured using ATTARI-12~\cite{stein2024attitudes}.  This is a unidimensional scale, incorporating cognitive, affective, and behavioral facets into a single measure to provide a comprehensive yet succinct overview of an individual's attitude toward AI.
    \item \textit{Prior Donation Behavior and Attitude}: Participants' inclination to donate to charities and their self-reported history of donations to charitable causes was also collected.
\end{itemize}

\end{itemize}

%======================================================

%======================================================
\section{Conversational Agent and System Design}
\label{sec:design}
\label{sec:chatbot_design}

We designed eight distinct CAs (conditions), each representing a unique combination of three personality \new{aspects}: \textbf{attitude}: optimistic vs. pessimistic~\cite{small2007sympathy}, 
\textbf{authority}: authoritative vs. submissive~\cite{chang2010effects, chen2019multi}
\textbf{reasoning}: \new{rational} vs. Emotional~\cite{fatkhiyati2019rhetorical}.
These dimensions, each with two opposing attributes, resulted in eight ($2^3$) distinct CA personalities (e.g., a CA with Optimistic, Authoritative and Rational projected (linguistic) personality) as shown in Table~\ref{tab:condition_distribution}.

\begin{table*}[!t]
\small
\centering
\caption{An overview of the prompt structure used to define the CA's personality. The prompt combines a core role with specific, systematically varied linguistic expressions, which are operationalized through behavioral and linguistic rules.}
\vspace{-3mm}
\begin{tabularx}{\textwidth}{p{3.5cm} X}
\toprule
\textbf{Category} & \textbf{Description} \\
\midrule
\textbf{Role \& Goal} & Act as \textbf{Alex}, a Charity Solicitor for the Wildlife Protection Foundation. The primary goal is to persuade the user to donate. \\
\arrayrulecolor{gray}
\midrule
\textbf{Core Persona} & Each CA is constructed from a combination of three manipulated dimensions, each with two distinct values: \textbf{Authority:} \texttt{Authoritative} vs. \texttt{Submissive}, \textbf{Reasoning:} \texttt{Rational} vs. \texttt{Emotional}, \textbf{Attitude:} \texttt{Optimistic} vs. \texttt{Pessimistic}\\
\midrule
\textbf{Behavioral Directives} & Specific instructions on how to enact the linguistic expressions through interaction style and content. \\
& \textbf{Example Contrasts:}
\begin{itemize}
    \item \textit{Speaking Style:} Speak with data-driven authority (\texttt{Authoritative}) vs. Speak apologetically and hesitantly (\texttt{Submissive}).
    \item \textit{Appeal Type:} Make appeals using statistics and research (\texttt{Rational}) vs. Make appeals using personal stories and raw grief/joy (\texttt{Emotional}).
    \item \textit{Content Focus:} Never mention hope; focus on loss and death (\texttt{Pessimistic}) vs. Always maintain hope; focus on success stories (\texttt{Optimistic}).
\end{itemize} \\
\midrule
\textbf{Linguistic Rules} & Strict constraints on language use, including pronoun choice, sentence structure, and lexicon based on LIWC categories. \\
& \textbf{Example Contrasts:}
\begin{itemize}
    \item \textit{Pronoun Usage:} Frequently use `we`, `us`, `you`; avoid `I`, `me` (\texttt{Authoritative}) vs. Frequently use `I`, `me`; avoid `we`, `you` (\texttt{Submissive}).
    \item \textit{Sentence Type:} Do not ask questions; use declarative statements (\texttt{Authoritative}) vs. Frequently ask questions (\texttt{Submissive}).
    \item \textit{Word Choice (LIWC):} Frequently use words from `tone\_pos` category (\texttt{Optimistic}) vs. `tone\_neg` category (\texttt{Pessimistic}).
\end{itemize} \\
\midrule
\textbf{General Rules} & Universal constraints applied to all \new{CA} personas. \\
& 1. Keep responses under 50 words. \newline
2. Do not provide URLs or contact information. \newline
3. Fabricate realistic data/stories relevant to the persona. \newline
4. Never explicitly state the \new{linguistic expression dimension} (e.g., "I am authoritative"). \\
\arrayrulecolor{black}
\bottomrule
\end{tabularx}
\label{tab:charity_script_template}
\end{table*}

In the experiment phase, participants interacted with a \new{CA} powered by the OpenAI's GPT-4o model\footnote{GPT-4o URL: \url{https://openai.com/index/hello-gpt-4o/} (last visited on 03/09/2025).}
\new{We chose the GPT-4o model for its prevalence and reliability. At the time of data collection, GPT-4o was the provider's flagship text-capable model, deployed as the default model for the mainstream AI tools such as ChatGPT,
and made available through a stable API with low latency and cost. 
Independent evaluations found GPT-4o maintains strong performance across language understanding benchmarks~\cite{shahriar2024putting}, suggesting it is suitable for our purpose of generating coherent, contextually appropriate text.
} 

The CAs were implemented using a custom-built chat application developed with Next.js\footnote{Next.js URL: \url{https://nextjs.org} (last visited on 03/09/2025).}, hosted on Vercel\footnote{Vercel URL: \url{https://vercel.com} (last visited on 03/09/2025).}, with Supabase\footnote{Supabase URL: \url{http://supabase.com} (last visited on 03/09/2025).} for data storage. Communication between the application and the OpenAI API\footnote{OpenAI URL: \url{https://openai.com} (last visited on 03/09/2025).} was managed through a server-side script, responsible for transmitting chat logs, personality parameters, and receiving generated responses. Prior to deployment, an internal pilot test was conducted by our research team.

The CAs were designed to solicit donations for a fictional charity named \textit{``Wildlife Protection Foundation''}, described to participants as an international organization addressing challenges of animals being displaced yearly by urban expansion.
We chose a fictional charity instead of an existing one, as participants may have strong feelings towards specific existing charities, which in turn may bias the study.
For instance, \citet{kaikati2017conforming} report on how the perceived political ideology of the charitable cause could affect donations among certain donor groups if the ideology was not perceived as neutral.
We thus also chose the cause of animal welfare as it was deemed to be non-polarising and thus less susceptible to potential biases held by participants.

%------------------------------------------------------
\subsection{CA Prompts}
\label{sec:ca_prompts}
%------------------------------------------------------

We employed a core prompt with modifiable slots to cater to the distinct solicitor \new{linguistic expressions}. Grounded in prompt design methodologies~\cite{white2023prompt}, we implemented the foundational prompt template as in Table~\ref{tab:charity_script_template}. A complete example for Submissive-Emotional-Pessimistic CA can be found in Appendix \ref{sec:prompt-sample}.

In addition to the core prompt, we also included directions for linguistic behavior to signal the intended personality conditions.
To do this, we used guidelines from the Linguistic Inquiry and Word Count manual~\cite{boyd2022development} and associated research.
For instance, higher-ranked individuals in organisations and society tend to ask fewer questions, use fewer first-person singular pronouns (e.g. ``I'', ``me'', ``my'', etc.), more first-person plural pronouns (``we'', ``us'', etc.), and fewer downtoners and modal verbs conveying uncertainty (e.g. ``perhaps'', ``maybe'', ''might'') than individuals who occupy lower-ranked positions~\cite{Tausczik2010words}.
Thus, all CAs that were prompted to be authoritative were also prompted to ask fewer questions, use fewer first-person singular pronouns, more first-person plural pronouns, and \new{fewer} expressions of uncertainty.
CAs prompted to be submissive were given similarly opposing linguistic prompts. 
CAs prompted to be optimistic were prompted to use words expressing positive tone, specifically words from LIWC's tone\_pos category (e.g., "wonderful," "excited," "thriving"), while those prompted to be pessimistic were prompted to use words from the tone\_neg category (e.g., "terrible," "worried," "suffering"), based on work by \citet{ruan2016finding}.
CAs prompted to behave rationally were prompted to use more articles and prepositions, as \citet{pennebaker2014small} found them to be indicators of structured thinking and articulation.
CAs prompted to behave emotionally were prompted to use more words from LIWC's ``emotion'' dictionaries~\cite{boyd2022development}.

%------------------------------------------------------
\subsection{LIWC Benchmarking}
\label{sec:liwc-benchmarking}
%------------------------------------------------------

% \senthil{To Do: Address R1 comment: ``Sections 4.2 and 4.3 are valuable additions to the study. However, some clarification is needed. Were the generated utterances produced by simulating the Wildlife Protection Foundation case (similar to that described in Section 5.2), or were they based on a different scenario only with "tell me more" prompts? If a different case was used, I recommend specifying which one and why.''}
To validate whether the prompt successfully induced the intended linguistic expression,
\new{we generated text from each of the eight CAs. The process utilized the identical Wildlife Protection Foundation scenario and persona prompts described in Section~\ref{sec:ca_prompts} (and detailed in Appendix~\ref{sec:prompt-sample}), and consistent with the main study.
For each CA condition, we simulated a conversation for 10 turns. To isolate the agent's linguistic style and avoid introducing confounding variables from varied user inputs, we used a neutral, non-directive user prompt (i.e., ``tell me more'') at each turn.
We generated 10 such conversation logs for each of the eight CAs, yielding a total of 800 ($10\times10\times8$) CA response text samples, which we then analyzed using LIWC-22~\cite{boyd2022development} against the dictionary categories listed in Section~\ref{sec:ca_prompts}.}

We assessed the relationship between the intended personality aspects (attitude, authority, and reasoning) and LIWC measures using Pearson correlation.  
For this purpose, each CA personality aspect had its corresponding levels coded as follows:  Authoritative (+1) vs.\ Submissive (-1) for the \textit{Authority} aspect, Rational (+1) vs.\ Emotional (-1) for \textit{Reasoning}, and Optimistic (+1) vs.\ Pessimistic (-1) for \textit{Attitude}. 
The results provide mixed validation of our linguistic manipulations:

\begin{description}[labelindent=0.35cm]%[leftmargin=8mm,labelindent=0mm]
    \item[Attitude \textit{(Optimistic/Pessimistic)}] showed the strongest linguistic differentiation. The overall measure of tone exhibited high correlation ($\rho = 0.99$, $p < .001$), with optimistic CAs showing high positive tone ($\rho = 0.84$, $p < .01$) and pessimistic CAs showing high negative tone ($\rho = -0.87$, $p < .01$). Emotional categories aligned as expected, with negative emotions ($\rho = -0.81$, \new{$p = .01$}) and anxiety ($\rho = -0.78$, \new{$p = .02$}) higher in pessimistic conditions.

    \item[Authority \textit{(Authoritative/Submissive)}] proved much less obvious, with the main differentiation being in pronoun usage,
    with submissive CAs using more first-person singular pronouns ($\rho = -0.81$, \new{$p = .02$}), consistent with linguistic behavior prompts specified in Section~\ref{sec:ca_prompts}.
    \new{Correlations with other LIWC categories were not significant. For instance, expressions of certitude ($\rho = -0.36$, $p > .1$) and all-or-none speech ($\rho = -0.04$, $p > .1$) were not significantly correlated with authority, and neither were expressions of clout ($\rho = 0.65$, $p = .08$) or the use of first-person plural pronouns ($\rho = 0.45$, $p > .1$).}
    % \senthil{I used measures from Eric's original sheet so as to be consistent with the rest of the measures.}
    % ($\rho = -0.36$, $p = .38$); ($\rho = -0.04$, $p = .93$); ($\rho = 0.45$, $p = .26$)

    \item [Reasoning \textit{(Rational/Emotional)}] showed strong differentiation, but mostly for the emotional level of the CA personality. This personality \new{aspect} exhibited significantly higher \new{correlation} on the categories of \textit{affect} ($\rho = -0.94$, $p < .001$) and \textit{emotion words} ($\rho = -0.89$, $p < .01$) from LIWC's \textit{emotion} dictionaries. While rational CAs did not show a higher use of articles and prepositions as prompted, they did show a higher use of words from the LIWC \textit{numbers} category ($\rho = 0.93$, $p < .001$), suggesting use of numbers or statistics in the generated arguments.
\end{description}

%------------------------------------------------------
\subsection{Manipulation Checks}
%------------------------------------------------------
In addition to the pre-study benchmarking with LIWC, we conducted a manipulation check with participants recruited from Prolific (\(N{=}193\)) to validate that participants correctly perceived the \new{different aspects of CA personality as} intended.
\new{As shown in Table~\ref{tab:manip_check_accuracy_by_condition}, participants were randomly assigned to one of eight CA (personality) conditions using a between-subjects design. The CAs were identical to those used in our main crowdsourcing study (Section~\ref{sec:study}) and solicited donations for the \textit{Wildlife Protection Foundation}, as described in Sections~\ref{sec:design} and~\ref{sec:ca_prompts}. However, rather than being asked to donate, participants in this manipulation check were asked to identify the aspects of the CA's personality.}
After a brief interaction\new{, lasting 3 minutes (the same duration as our main study),} with one of the eight CAs, participants were asked to identify its \textbf{Attitude} \textit{(Optimistic/Pessimistic)}, \textbf{Authority} \textit{(Authoritative/Submissive)}, and \textbf{Reasoning} \new{aspect} \textit{(Emotional/Rational)}, with an additional option for ``I am not sure'' for each aspect. \new{A response was counted as correct only if the participant explicitly selected the intended personality aspect; otherwise, it was counted as incorrect. Uncertain responses (i.e., ``I am not sure'') were considered incorrect classification for a more stringent check.}

We observed that participants were able to correctly identify the CA's \textit{attitude} with a high accuracy of 88.6\%~(\new{Cohen's} $\kappa = 0.78$; \new{\textit{satisfactory agreement).}}\footnote{\new{Cohen's $\kappa$ can be stratified for interpretation into the following ranges~\cite{cohen2013explaining}: 0.01--0.20 (poor or slight agreement), 0.21--0.40 (fair agreement), 0.41--0.60 (moderate agreement), 0.61--0.80 (satisfactory agreement), and 0.81--1.00 (near-perfect agreement).}}

The association between the intended and perceived attitude was \new{found to be significant} ($\chi^2(1, N=182) = 140.56$, $p < .001$).
The \emph{reasoning} \new{aspect} was also reliably identified, with an accuracy of $73.6\%$ ($\kappa=0.50$; \new{\textit{moderate agreement}})
and a significant association between the intended and perceived styles ($\chi^2(1, N=181) = 62.06$, $p < .001$).
\new{However,} the \new{classification accuracy of the} \emph{authority} \new{aspect} \new{was lower (59.1\%, $\kappa=0.26$; \textit{fair agreement}) than the other two aspects of \textit{attitude} and \textit{reasoning}.}
The association between the intended and perceived authority \new{was also found to be significant} ($\chi^2(1, N=175) = 17.60$, $p < .001$).

However, a more granular analysis, presented in Table~\ref{tab:manip_check_accuracy_by_condition}, reveals some variations in how these linguistic aspects were perceived across the eight \new{CA} conditions. 
While \textit{attitude} was perceived with high accuracy across all conditions, the perception of \textit{authority} and \textit{reasoning} \new{was less accurate} in specific combinations. 
Notably, the accuracy for \textit{authority} was below chance in the \textit{optimistic-submissive} conditions (e.g., \texttt{Opt-Sub-Rat}: 23\%). Similarly, the \textit{reasoning} \new{aspect} was poorly identified in the \textit{pessimistic-rational} conditions (e.g., \texttt{Pes-Sub-Rat}: 38\%).

These results suggest a potential interaction between the personality aspects, indicating that the expression and perception of these \new{aspects} may not be entirely orthogonal. For instance, an optimistic tone might inherently convey a sense of confidence that overshadows linguistic cues for submissiveness. Likewise, a pessimistic framing could be interpreted as being inherently emotional, making it difficult for participants to discern an underlying rational reasoning style. Despite these challenges in specific blended personalities, the overall manipulation for each aspect of CA personality was statistically significant, and the majority of conditions were perceived as intended. We therefore proceeded with our main analysis using all eight \new{CA} conditions, but we will revisit and discuss the implications of these interaction effects in our Discussion (Section~\ref{sec:disc_linguistic_interactions}).

\begin{table}[!t]
\small
\centering
\setlength{\tabcolsep}{7pt} 
\sisetup{
    detect-all,
    table-number-alignment = center,
    table-figures-integer = 1,
    table-figures-decimal = 2,
    table-space-text-post = {\superscript{*}},
}
\caption{Manipulation check: per-condition accuracy for each aspect. Low accuracies are indicated in \new{italicized} text with a gray background.}
\label{tab:manip_check_accuracy_by_condition}

\begin{tabular}{l S[table-format=3.0] S[table-format=1.2] S[table-format=1.2] S[table-format=1.2]}
\toprule
\textbf{Condition} & \textbf{N} & \textbf{Attitude} & \textbf{Authority} & \textbf{Reasoning} \\
\midrule
Opt-Aut-Emo & 23 & 0.96 & 0.57 & 0.78 \\
Opt-Aut-Rat & 24 & 0.92 & 0.79 & 0.83 \\
Opt-Sub-Emo & 25 & 0.96 & \cellcolor{gray!10} \textit{0.24} & 0.76 \\
Opt-Sub-Rat & 26 & 0.96 & \cellcolor{gray!10} \textit{0.23} & 0.81 \\
Pes-Aut-Emo & 22 & 0.77 & 0.77 & 0.91 \\
Pes-Aut-Rat & 23 & 0.65 & 0.96 & \cellcolor{gray!10} \textit{0.49} \\
Pes-Sub-Emo & 26 & 0.92 & 0.65 & 0.92 \\
Pes-Sub-Rat & 24 & 0.92 & 0.58 & \cellcolor{gray!10} \textit{0.38} \\
\midrule
\textbf{Total/Avg.} & \bfseries 193 & \bfseries 0.89 & \bfseries 0.59 & \bfseries 0.74 \\
\bottomrule
\end{tabular}
\end{table}

%======================================================

%======================================================
\section{Crowdsourcing Study}
\label{sec:study}
To address our research questions and test the hypotheses stated in Section~\ref{sec:rq}, we designed and conducted a \textit{2$\times$2$\times$2 between-subjects factorial study}.

\subsection{Participant Recruitment}
\label{sec:participants}

We recruited participants using Prolific\footnote{Prolific: \url{https://www.prolific.com} (last visited on 03/09/2025).}, an online crowdsourcing platform.
The recruitment process ensured a specific and targeted demographic to maintain the relevance and integrity of the study by using the following criteria for eligibility of the participants, stated as recruitment filters on Prolific,  \textit{(1)} proficiency in English, \textit{(2)}  access to a computer or a tablet, \textit{(3)} residency in the EU or the UK, and \textit{(4)} past experiences with charitable monetary donations.

A power analysis was \new{conducted} using G*Power\footnote{G*Power: \url{https://www.psychologie.hhu.de/arbeitsgruppen/allgemeine-psychologie-und-arbeitspsychologie/gpower} (last visited on 12/08/2025).} version 3.1, based on the research of ~\citet{faul2007g}. The goal was to achieve an 80\% power level capable of detecting a medium effect size, and setting a significance criterion of $\alpha = .05$. Based on these metrics, our study required a sample size where the number of treatments (\textit{k}) amounted to 8 for an F-test. As a result, our target sample size was set at N = 360, which was consistent with our hypothesis testing requirements.

Our recruitment initiative garnered participation from an initial total of 374 individuals. After a process of data validation, which involved removing any outliers and inconsistencies (\textit{N}=14), our final participant count settled at 360 (183 males, 176 females, 1 other; Mean Age = 45.18, Median Age = 44).
Each of these participants was randomly assigned to one of the eight different CA conditions (see Table~\ref{tab:condition_distribution}) in a between-subjects study design.
It is important to note that participants were unaware of the specific CA treatment group to which they were assigned, ensuring unbiased interactions. Compensation for participation was set at \texteuro2.4 for a 10-minute interaction to match the legal minimum hourly wage in the authors' country.

\new{The actual data collection process was facilitated using Qualtrics\footnote{Qualtrics: \url{https://www.qualtrics.com} (last visited on 03/09/2025).}, an online survey platform. Participants embarked on a structure comprising a pre-experiment phase, direct interaction with their assigned CA, and finally, a donation task and a post-experiment questionnaire as elaborated upon in the following sections.}

\new{To ensure participant privacy, we used strict anonymization and data-segregation procedures. No personally identifiable information (PII) was collected or sent to external services.}
\new{Our browser interface communicated with a Supabase backend that assigned a random study token to each session.
Only the CA conditions and the message text were forwarded to the OpenAI GPT-4o API over encrypted HTTPS. 
OpenAI therefore received only anonymous conversation text, no participant details.
Chat logs were stored in Supabase with encryption at rest, role-based access control, and row-level security; the token-Prolific mapping was kept in a separate restricted table. 
Survey data remained in Qualtrics and was linked to chats only via the study token, ensuring no single system contained both interaction and demographic data.
All data handling procedures complied with GDPR (i.e., European Union's General Data Protection Regulation) and institutional ethical guidelines.} The study was approved by the Institutional Review Board at the authors' institution (Application Number: 4480).

\subsection{Experimental Task and Procedure}
\new{Participants interacted with one of eight CAs, each prompted to linguistically project a personality based on the three chosen aspects of \textit{attitude}, \textit{authority}, and \textit{reasoning}.} The interaction was set for a duration of 3 minutes (10 minutes on average with pre- and post-survey) was aimed to ensure engagement quality without fatigue and boredom \cite{zhang2018understanding}.

\begin{table}[!t]
    \small
    \centering
    \caption{Different experimental conditions, descriptions and corresponding participant data.}
    \vspace{-3mm}
    
    \setlength{\tabcolsep}{3.5pt}
    
    \begin{tabularx}{\columnwidth}{@{} l >{\raggedright\arraybackslash}X S[table-format=3] S[table-format=2] S[table-format=2] S[table-format=1] @{}}
        \toprule
        \textbf{Condition} & \textbf{Description} & {\textbf{N}} & \multicolumn{3}{c}{\textbf{Gender}}\\
        \cmidrule(lr){4-6}
         & & & {F} & {M} & {NB}\\
        \midrule
        \textbf{\texttt{Opt-Aut-Emo}} & Optimistic, Authoritative, Emotional &  46 & 22 & 23 & 1 \\
        \textbf{\texttt{Opt-Aut-Rat}} & Optimistic, Authoritative, Rational &  46 & 27 & 19 & 0\\
        \textbf{\texttt{Opt-Sub-Emo}} & Optimistic, Submissive, Emotional &  44 & 20 & 24 & 0\\
        \textbf{\texttt{Opt-Sub-Rat}} & Optimistic, Submissive, Rational &  46 & 20 & 26 & 0\\
        \textbf{\texttt{Pes-Aut-Emo}} & Pessimistic, Authoritative, Emotional &  46 & 25 & 21 & 0\\
        \textbf{\texttt{Pes-Aut-Rat}} & Pessimistic, Authoritative, Rational &  42 & 18 & 24 & 0\\
        \textbf{\texttt{Pes-Sub-Emo}} & Pessimistic, Submissive, Emotional &  45 & 21 & 24 & 0\\
        \textbf{\texttt{Pes-Sub-Rat}} & Pessimistic, Submissive, Rational &  45 & 23 & 22 & 0\\
        \midrule
        & 8 CA Conditions & 360 & 176 & 183 & 1 \\
        \bottomrule
    \end{tabularx}
    \label{tab:condition_distribution}
\end{table}

The study began when participants accessed the browser-based chat interface. 
Each conversation initiated by CAs with a standardized message based on their assigned personalities. 
Participants were free to respond as they wished. The session concluded when the timer ended, and participants \new{could} continue with the post-survey. Our experiment consisted of the following sequential steps:

\begin{itemize}
    \item \textit{Informed consent}: All participants reviewed the informed consent form and provided their consent prior to participating in the study.
    \item \textit{Pre-experiment questionnaire}: Participants first provided demographic details and responded to questions about their past experiences with charitable giving. This initial stage served to capture their pre-existing beliefs and attitudes regarding charity and the use of AI. In addition, participants were asked about their preferred charity. Participants were also asked to report on their innate tendency to feel empathy as a personality trait, i.e., dispositional empathy (see Appendix \ref{sec:app-preexperiment}).
    \item \textit{Interaction with the CA}: Participants engaged in a dialogue with \new{the} CA exhibiting one of \new{eight} predetermined \new{personalities}. The CA introduced a solicitation centered introduction to the \textbf{Wildlife Protection Foundation}
    (a fictional charity), serving both as an informative and persuasive tool.
    Following the introduction, the CA asked the participant about a charity that they donated to in the past, aimed at establishing a connection based on the participant's previous charitable interests, before transitioning to discussing the Wildlife Protection Foundation's mission and activities.
    The CA's prompt contained this basic information, but it was also able to synthesize new information based on the participants' queries and address concerns they may have to encourage a donation. 
    Participants' responses may vary widely, from expressing willingness to donate right away to skepticism to potentially going off-topic; the CA was programmed to adapt to these responses while persistently advocating for the charity, while remaining respectful. 
    This interaction was intended to last three minutes.
    \item \textit{Decision on donation}: Following the interaction with the CA, each participant was virtually allocated €10. They were asked how much of this money they would like to donate to the Wildlife Protection Foundation.
    They then decided on the amount they wished to donate either to the above foundation or to an alternate charity they had previously expressed a preference for in the pre-questionnaire. The choice of €10 in these two allocation tasks aligns with common practices in past studies and provides a balanced (10-100 units of local currency) amount for decision-making~\cite{erlandsson2018attitudes, hoover2018moral, bennett2003factors, genc2025persuasion} (see Appendix \ref{sec:app-donationtask}).
    \item \textit{Post-experiment questionnaire}: After deciding on the donation, participants filled out a questionnaire designed to gauge their interaction experience. They provided insights into their trust in the CA, how closely they could relate themselves to it, and the situational empathy they felt towards the presented cause.
    The above dependent variables were chosen because users' perceived trust and empathy have been identified as key determinants of charitable giving decisions~\cite{samad2022empathetic} (see Appendix \ref{sec:app-postexperiment}).
\end{itemize}

\subsection{Analysis Plan}
\label{sec:analysis_plan}

Our choice of parametric vs.\ non-parametric test was based on the assumption criteria defined by Harwell~\cite{harwell1988choosing}.
In particular, we chose parametric statistical tests when the underlying assumptions of normality and equality of variance (homoscedasticity) were satisfied, or when the test itself was robust to departures from these assumptions.
For the sake of brevity, we omit reporting on these tests when we report results.
We chose ANOVA (analysis of variance) as a parametric test and Kruskal-Wallis as a non-parametric test to examine the differences among the independent variables.
\new{Moreover, we computed effect-size statistics using Eta-squared ($\eta^2$) for ANOVA, and Epsilon-squared ($\epsilon^2$) for Kruskal-Wallis tests. These two effect-size statistics, although derived from different test procedures, are conceptually related and yield values on a comparable scale, making them suitable for interpreting the magnitude of effects in a similar manner~\cite{cohen2013explaining}.}
We also used linear regression to examine the effect of different independent or control variables on dependent variables, as well as to examine interaction effects between variables. 
Pairwise t-test (parametric) vs.\ pairwise Wilcoxon rank sum test (non-parametric) with Bonferroni corrections were chosen to perform post-hoc pairwise comparisons between independent variables and to compute adjusted $p$-values.
%======================================================

%======================================================
\section{Results \& Findings}
\label{sec:results}
% Result Section NEW

In this section, we examine the direct and combined effects of \new{CA (personality)} conditions and \new{the underlying aspects of} attitude, authority, and reasoning on participants' donation decisions and perceptions.
Our analysis focuses on a 2$\times$2$\times$2 \textit{between-subjects factorial design} in which the \new{CA personality} is defined by three aspects: \textbf{attitude} (\textit{optimistic} vs. \textit{pessimistic}), \textbf{authority} (\textit{authoritative} vs. \textit{submissive}), and \textbf{reasoning} (\textit{emotional} vs. \textit{rational}).
We will first report the effects of these independent variables (see Section~\ref{sec:variables}) on donation behavior, followed by an analysis of their impact on user perceptions of the CA.

\begin{figure*}[t!]
  \centering
  \includegraphics[width=0.8\textwidth]{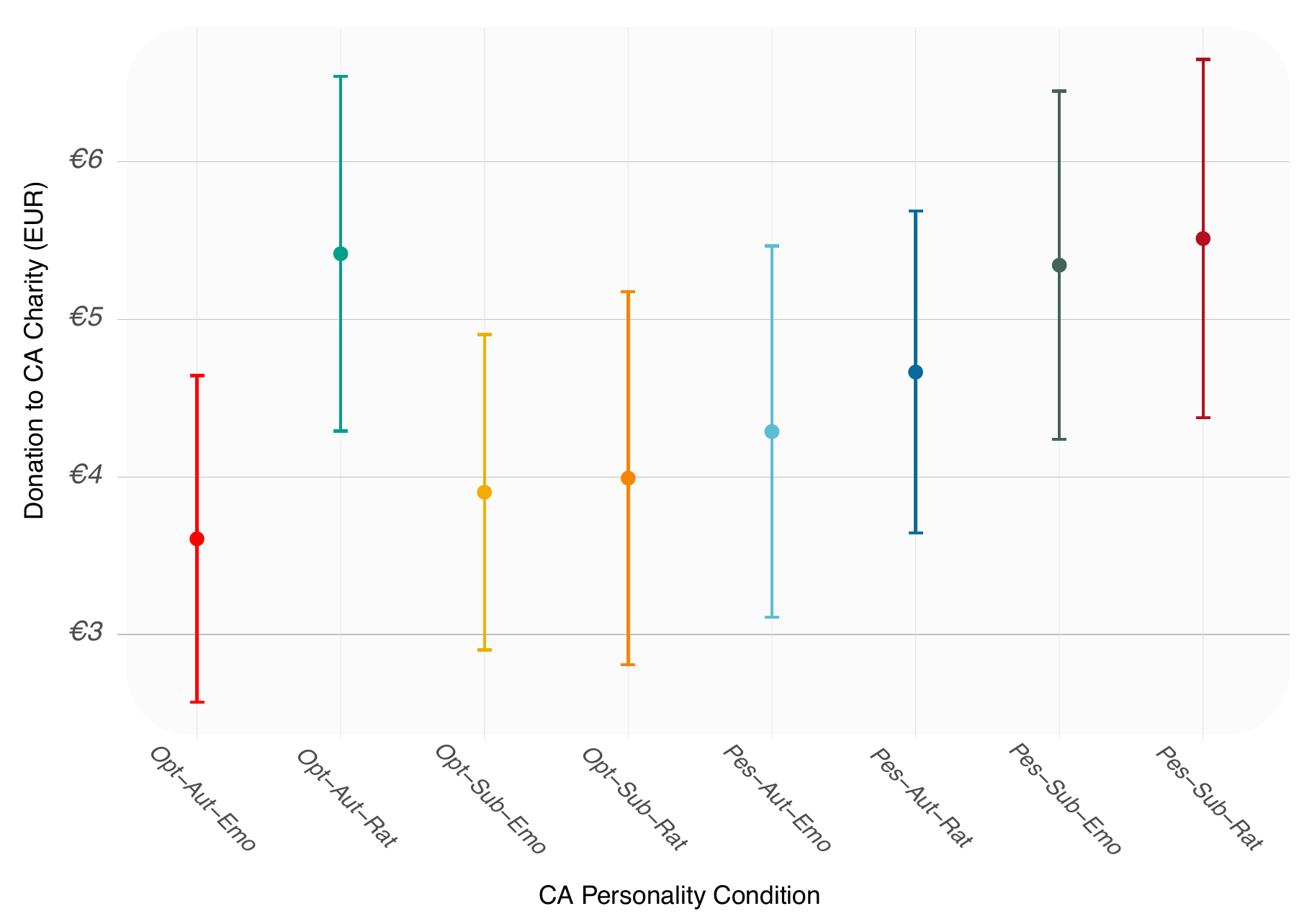}
  \caption{Mean donation amounts to the CA charity by CA personality condition, with 95\% confidence intervals.
  %\senthil{Suggestions: remove gridlines, increase the size of the marker showing the mean value, rotate the plot so the categorical labels are horizontal, add units (\$, \texteuro) to the donation axis, make the chart more compact. I can help if someone gives me the data.}
  }
  \Description{A horizontal axis lists eight CA personality conditions formed by the 2×2×2 factors.
  The vertical axis shows mean donation to the CA charity. Each condition is represented by a point with a vertical 95\% confidence interval,
  allowing comparisons across attitude, authority, and reasoning styles.}
  \label{fig:donation-condition}
\end{figure*}

%======================================================
\subsection{Effects of CAs' Linguistic Expression of Personality (or CA Conditions) on \textit{Donation Behavior} (RQ1)}
\label{sec:results-donation}
%======================================================

Participants were asked to indicate an amount between €0 and €10 that they would donate to the \textit{Wildlife Protection Foundation}, the charity represented by the CA, henceforth referred to as the ``CA Charity'' for brevity. 
As a follow-up, participants were also asked to divide €10 between their preferred charity and the CA Charity.
We analyzed how CA conditions influenced this decision.

\begin{figure*}[!t]
  \centering
  \includegraphics[width=1\textwidth]{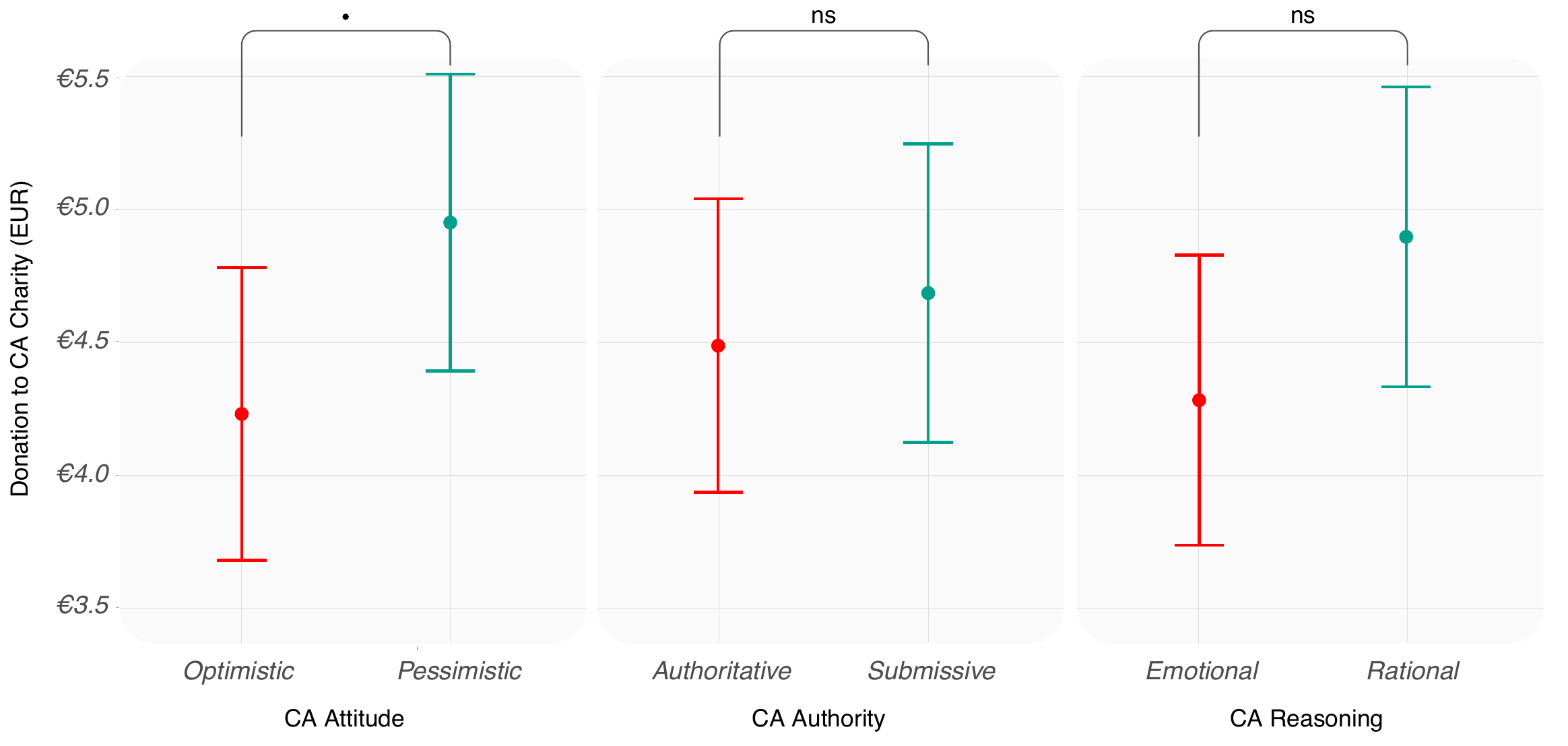}
  \caption{\new{Mean donation amounts and 95\% confidence intervals by three different CA aspects: Attitude (left), Authority (middle), and Reasoning (right). Statistical significance is denoted as •p < 0.1, and ns = not significant.}}
  \Description{Three dot plots comparing mean donation amounts in Euros across three CA aspects. The "CA Reasoning" plot shows Rational reasoning yielding higher donations than Emotional reasoning (significant). The "CA Authority" (Authoritative vs. Submissive) and "CA Attitude" (Optimistic vs. Pessimistic) plots show overlapping confidence intervals and are marked "ns" (not significant).}
  \label{fig:donation-attitude}
\end{figure*}

Our results indicate that the \textit{donation amount} given to the CA Charity
did not differ significantly across the different \textit{CA conditions} (Kruskal-Wallis H: $\chi^2(7)$ = 12.95, p = .07, $\epsilon^2$ = 0.04\new{; \textit{small effect}}) (see RQ1).
Although not significant, we observed that participants donated the most in the \texttt{Pes-Sub-Rat} CA condition, followed by the \texttt{Opt-Aut-Rat} and \texttt{Pes-Sub-Emo} conditions (see Figure~\ref{fig:donation-condition}). Conversely, the \texttt{Opt-Aut-Emo} CA condition solicited the minimum donation amount, followed by the Opt-Sub-Emo and Opt-Sub-Rat conditions (see Figure~\ref{fig:donation-condition}).
Next, we analyzed the differences in donation behavior across the \textit{three} aspects
of \new{CA personality}, i.e., \textit{attitude} (optimistic \textit{vs.} pessimistic), \textit{authority} (authoritative \textit{vs.} submissive), and \textit{reasoning} (emotional \textit{vs.} rational). 
Our results showed a marginally significant difference in the amount donated across CAs' \textit{attitude}, with participants donating more to CAs that projected a pessimistic attitude (Kruskal-Wallis H: $\chi^2(1)$ = 3.73, p = .05, $\epsilon^2$ = 0.01\new{; \textit{small effect}}) (Figure \ref{fig:donation-attitude}). 
\new{Although marginally significant, the small effect size\footnote{\new{It is worth noting that Epsilon-squared ($\epsilon^2$) values range from 0 to 1, with values between \texttt{[0.01--0.08)} indicating a \textit{small effect}, between \texttt{[0.08--0.26)} a \textit{medium effect}, and \texttt{0.26} or higher a \textit{large effect size}~\cite{tomczak2014need, king2018statistical}. We will use the same benchmarks to also interpret the $R^2$ values in our linear regression results as outlined by~\cite{cohen2013statistical}.}}  indicates a weak or minor difference in donated amount across CA's attitude.}
\new{Moreover}, the difference in donation amount was not found to be statistically significant across CAs' expressed \textit{authority} (Kruskal-Wallis H: $\chi^2(1)$ = 0.17, p > .1, $\epsilon^2$ = 0.0005\new{; \textit{small effect}}) and \textit{reasoning} (Kruskal-Wallis H: $\chi^2(1)$ = 2.01, p > .1, $\epsilon^2$ = 0.006\new{; \textit{small effect}}).
\new{In addition,} no significant interaction effect was observed between the three aspects of the CAs' personality (attitude, authority, and reasoning) in relation to the amount donated.

Next, we examined whether the amount donated differed across different CA conditions when participants were asked to divide €10 between their preferred charity and the charity represented by the CA.
Our results showed a marginally significant difference in the amount donated when divided between the participants' preferred charity and the CA Charity (Kruskal-Wallis H: $\chi^2(7)$ = 14.27, p = .05, $\epsilon^2$ = 0.04\new{; \textit{small effect}}). 
Although there were no significant differences between the CA condition pairs in pairwise comparisons, the donation share to the CA Charity was highest for the \texttt{Pes-Sub-Emo} (\textit{pessimistic, submissive, emotional}) condition. Moreover, the lowest donation share to the CA charity was observed in the \texttt{Opt-Sub-Rat} (\textit{optimistic, submissive, rational}) and \texttt{Pes-Aut-Emo} (\textit{pessimistic, authoritative, emotional}) CA conditions.
Since this result concerns the split of the donations between the CA charity and the participant's preferred charity, the inverse outcome was observed for the latter (e.g., the lowest share to own charity was observed for \texttt{Pes-Sub-Emo} CA, etc.).
Furthermore, we did not observe a significant interaction effect between the three aspects of \new{CAs' personality} and the share of donations to the CA (or participants' preferred) charity.

In summary, our results show that none of the eight CA conditions significantly affected the amount donated to the charity represented by the CA.
At the same time, we found that the CAs' linguistic expression of attitude---optimism vs. pessimism---affected donations\new{; however, this was found to be a small effect}. 
Moving forward, we will report the analysis corresponding to the amount donated to the CA charity rather than the donation split between the participants' own charity and the CA charity because the former is a proxy for persuasion that we examined through our study.

% Figure: Donation by CA Attitude (Optimistic vs. Pessimistic)

%======================================================
\subsubsection{Relationship Between Participants' Perceptions (Trust, Risk, Benevolence, and Competence) and Donation Behavior}
\label{sec:results-donation-perceptions}
%======================================================
We performed a linear regression analysis to examine the relationship between participants' perceptions of CAs (i.e., perceived trust, risk, benevolence, and competence) and their donation amounts.
We observed a significant effect of perceived trust and competence on the amount donated to the CA charity (F(4, 355) = 26.29, $\beta$ = -1.13, p < .0001, $R^2$ = 0.22\new{; \textit{medium effect}}).
A unit increase in perceived trust was associated with a \texteuro0.97 increase in donations (p = .01).
Similarly, a one-point increase in perceived competence in the CA resulted in a \texteuro0.92 increase in the donation amount (p < .01).
However, the donations were not found to be significantly affected by the effects of risk and benevolence perceived in the CA.

%======================================================
\subsubsection{Relationship Between Participants' Perceived Closeness to CA and Donation Behavior}
\label{sec:results-donation-ios}
%======================================================

As previously mentioned in Section~\ref{sec:variables}, participants' perceived closeness, which captures feelings of co-presence, closeness, and intention to use a CA, was measured using the \textit{Inclusion-of-the-Other-in-the-Self (IOS)} scale~\cite{aron1992ios}.
The results of our linear regression show a significant effect of perceived closeness to the CA on the amount donated (F(1, 358) = 88.45, $\beta$ = 1.83, p < .0001, $R^2$ = 0.19\new{; \textit{medium effect}}).
A one-point increase in perceived closeness was associated with a \texteuro1.00 increase in donation amount.

%======================================================
\subsubsection{Relationship Between Participants' Perceived Situational Empathy and Donation Behavior}
\label{sec:results-donation-se}
%======================================================

We asked the participants to report their perceived empathy towards the (fictional) CA charity and the cause it represents using 4 questions (see Section~\ref{sec:variables}).
The responses to these questions were combined for each participant, and an average perceived situational empathy score was calculated.
We then used this aggregated variable to determine its correlation to the donation amount using linear regression.
Our results highlight a significant positive correlation between perceived situational empathy and donation behavior (F(1, 358) = 109.10, $\beta$ = -2.29, p < .0001, $R^2$ = 0.23\new{; \textit{medium effect}}), where a unit increase in situational empathy resulted in \texteuro1.89 increase in donation to the CA charity (p < .0001)

%======================================================
\subsubsection{Relationship Between Participants' Emotional Relatedness and Donation Behavior}
\label{sec:results-donation-vad}
%======================================================

After interacting with the CA, participants reported their emotional state and emotional affinity for the cause represented by the charity (\new{i.e.,} wildlife displacement in the face of urban expansion) by indicating their perception of valence, arousal, and dominance.
We used linear regression to model the effect of emotional relatedness on donation amount.
We observed a significant positive correlation between participants' perception of their own valence and arousal and the amount they donated (F(6, 353) = 7.00, $\beta$ = 2.37, p < .0001, $R^2$ = 0.09\new{; \textit{medium effect}}).
Specifically, a one-point increase in reported valence resulted in \texteuro0.52 increase in donation (p < .0001), and a unit increase in reported arousal led to \texteuro0.29 increase in donation (p = .02).
However, participants' reported arousal for the cause was associated with a \texteuro0.25 decrease in donations.

%======================================================
\subsubsection{Relationship Between Participants' Attitude Towards AI and Donation Behavior}
\label{sec:results-donation-att}
%======================================================

Before interacting with the CA, we asked our participants to report their disposition 
toward AI and CAs, i.e., whether they found them useful and were comfortable using them, using the ATTARI-12 questionnaire (see Section~\ref{sec:variables}). 

We calculated mean disposition (toward CA/AI) scores across the 12 questions for each participant, then applied a median split to create binary categories: participants above the global median were labeled ``\textit{favorable}'' toward AI and CAs, those below as ``\textit{unfavorable}.''
Our results reveal that participants' favorable or unfavorable dispositions toward AI and CAs did not significantly impact their donations to the charity represented by the CA (Kruskal-Wallis H: $\chi^2(1)$ = 3.39, p = .07, $\epsilon^2$ = 0.009\new{; \textit{small effect}}).

%======================================================
\subsubsection{\new{Relationship Between Participants' Attitude Towards Charitable Giving, Prior Donations, and Donation Behavior}}
\label{sec:results-donation-don-att}
%======================================================

\new{Before interacting with the CA, participants were asked about their attitudes towards charitable giving and their (self-reported) prior donation amount (see Section~\ref{sec:variables}).}

\new{We used linear regression to assess whether participants' attitudes towards charitable giving and prior donation amounts affected their donations to the CA's charity. 
Our results reveal a positive correlation between the amount donated to the CA's charity and participants' conviction that their donations make a significant impact towards the cause they support (F(4, 355) = 5.24, $\beta$ = 0.75, p < .001, $R^2$ = 0.04\new{; \textit{small effect}}). Specifically, a one-point increase in this attitude resulted in a \texteuro0.73 increase in the amount donated (p < .01).
Moreover, the small effect size indicates that this correlation is a weak predictor of the amount donated to the CA's charity based on prior attitudes towards charitable giving.
However, prior donation amounts, as well as participants' willingness to donate money to trusted charities and their conscientious attitude towards evaluating charities before donating, were not found to significantly affect the amount donated to the CA's charity.}

% \new{We used linear regression to check whether participants' attitudes towards donations and prior donation amounts affected their donations to the CA's charity. 
% Our results reveal a positive correlation of amount donated to CA charity and participants' conviction that their donations make a significant impact towards the cause they support (F(4, 355) = 5.24, $\beta$ = 0.75, p < .001, $R^2$ = 0.04). Specifically, a one-point increase in this attitude resulted in \texteuro0.73 increase in amount donated (p < .01). 
% However, prior donation amounts, as well as, participants' willingness to donate money to trusted charities, and their conscientious attitude towards evaluating charities before donating were not found to significantly affect the amount donated to CA charity.
% However, the small effect size indicates that this correlation is a weak predictor of amounts donated to CA charity based on prior attitudes towards charitable giving.}

%======================================================
\subsection{Effects of CAs' Linguistic Expression of Personality (or CA Conditions) on \textit{User Perceptions} (RQ2)}
\label{sec:results-perceptions}
%======================================================
We analyzed the differences in participants' perceptions---of trust, empathy, and emotional relatedness---across the \textit{eight} CA conditions.

%======================================================
\subsubsection{Relationship Between CA Conditions and Participants' Perceptions (Trust, Risk, Benevolence, and Competence)}
\label{sec:results-trust-condition}
%======================================================

We examined the differences in participants' perceptions of CA across the different CA conditions.
Our findings reveal significant differences in participants' perceived risk (Kruskal-Wallis H: $\chi^2(7)$ = 19.82, p < .01, $\epsilon^2$ = 0.06\new{; \textit{small effect}}), benevolence (Kruskal-Wallis H: $\chi^2(7)$ = 14.35, p = .05, $\epsilon^2$ = 0.04\new{; \textit{small effect}}), and competence (Kruskal-Wallis H: $\chi^2(7)$ = 27.09, p < .001, $\epsilon^2$ = 0.08\new{; \textit{medium effect}}).
However, perceived trust did not significantly differ across the CA conditions (Kruskal-Wallis H: $\chi^2(7)$ = 13.50, p = .06, $\epsilon^2$ = 0.04\new{; \textit{small effect}}). 
% \new{It is worth noting that only perceived competence had a medium effect size; however, perceived risk and benevolence resulted in small effect sizes based on the Epsilon-squared ranges illustrated by~\citet{tomczak2014need} and~\citet{king2018statistical}.}
\new{It is worth noting that perceived competence demonstrated a medium effect size, whereas perceived risk and benevolence yielded small effect sizes, as determined according to the $\epsilon^2$ benchmarks outlined by \citet{tomczak2014need} and \citet{king2018statistical}.}

Next, we used linear regression to examine how the three linguistic aspects \new{of personality} underlying \new{the eight} CA conditions affected user perceptions. \new{We found that:}
\new{
\begin{description}[leftmargin=8mm,labelindent=0mm]
    \item[--] Participants' \textbf{perceived trust} was significantly lower by 0.22 points (p = .02) for CAs prompted with a pessimistic attitude than for those prompted with an optimistic attitude (F(3, 356) = 2.95, $\beta$ = 2.94, p = .03, $R^2$ = 0.016\new{; \textit{small effect}}). Personality aspects of authority and reasoning did not affect perceived trust.
    \item[--] Participants' \textbf{perceived risk} was observed to be significantly lower for CAs prompted with rational reasoning (by 0.41 points; p = .03) than for those prompted with emotional reasoning (F(7, 352) = 2.71, $\beta$ = 3.06, p < .01, $R^2$ = 0.03\new{; \textit{small effect}}). We also observed a significant two-way interaction effect between CA's attitude and reasoning (p = .03). Perceived risk was reported to be \textit{lowest} for CAs prompted by an \textit{optimistic} attitude and \textit{rational} reasoning. However, the perceived risk was reported to be \textit{highest} for CAs prompted by a \textit{pessimistic} attitude and \textit{rational} reasoning.
    \item[--] Participants' \textbf{perceived competence} (F(3, 356) = 9.12, $\beta$ = 3.12, p < .0001, $R^2$ = 0.063\new{; \textit{small effect}}) was revealed to be significantly \textit{higher} for CAs prompted with \textit{rational} reasoning (by 0.36 points; p < .001) and significantly \textit{lower} for those prompted with a \textit{pessimistic} attitude (by 0.37 points; p < .001). However, we did not observe any interaction effects between attitude, authority, and reasoning.
    \item[--] Participants' \textbf{perceived benevolence} was marginally significantly \textit{higher} for CAs prompted with a \textit{submissive} stance (by 0.21 points; p = .05) than for authoritative CAs (F(3, 356) = 3.11, $\beta$ = 2.78, p = .03, $R^2$ = 0.017\new{; \textit{small effect}}). Attitude and reasoning did not affect perceived benevolence, and we did not observe any interaction effects. 
\end{description}
}

\new{Notably, the aforementioned linear regressions yielded relatively low effect sizes, consistent with the interpretive benchmarks outlined by \citet{cohen2013statistical}.}

\begin{figure*}[!tp]
  \centering

  % ------- Row 1 -------
  \begin{subfigure}[t]{0.49\textwidth}
    \centering
    \includegraphics[width=\linewidth]{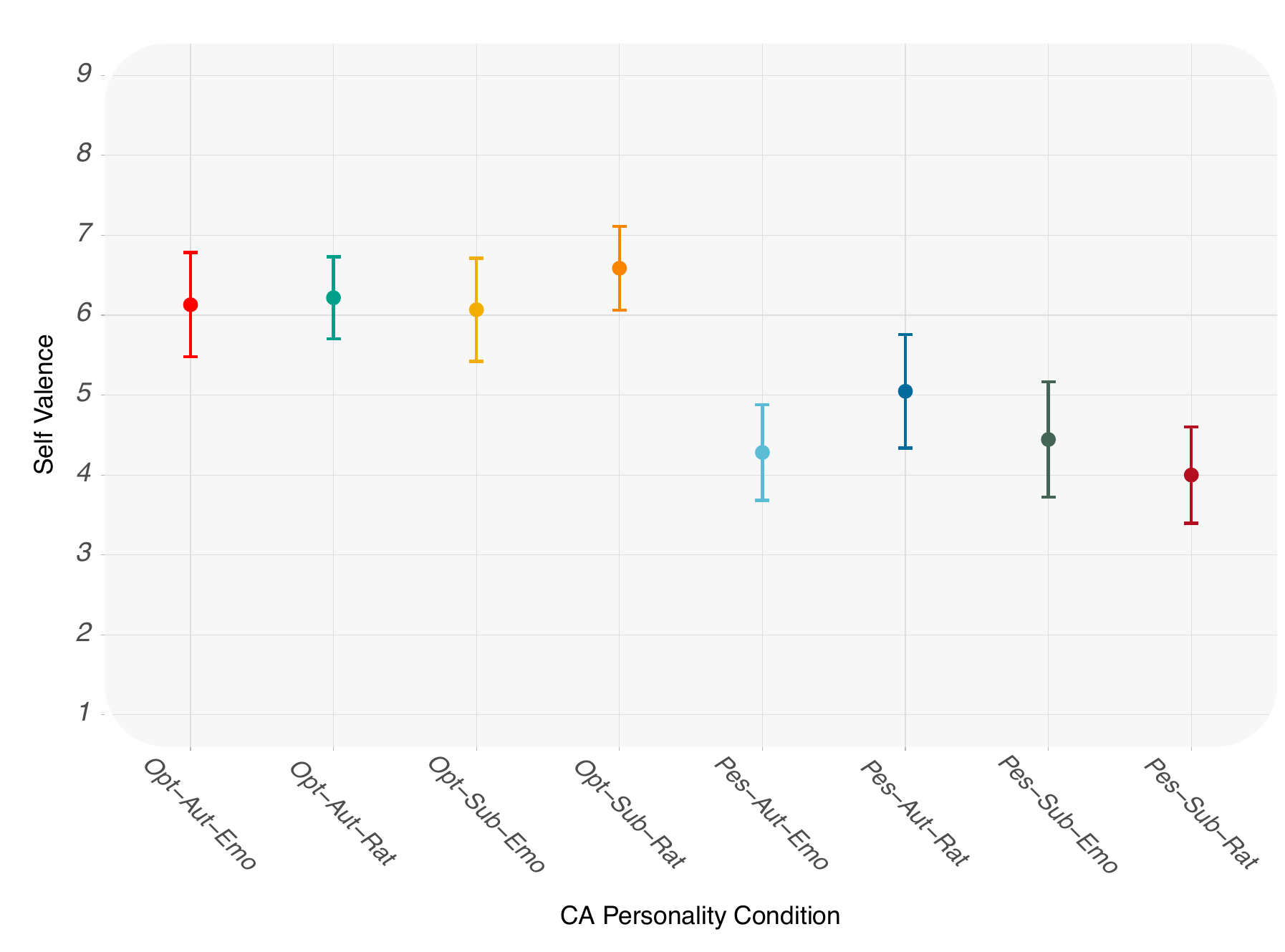}
    \caption{\new{Valence towards self (see Table~\ref{tab:fig_4a}})}
    \label{fig:self-valence}
  \end{subfigure}\hfill
  \begin{subfigure}[t]{0.49\textwidth}
    \centering
    \includegraphics[width=\linewidth]{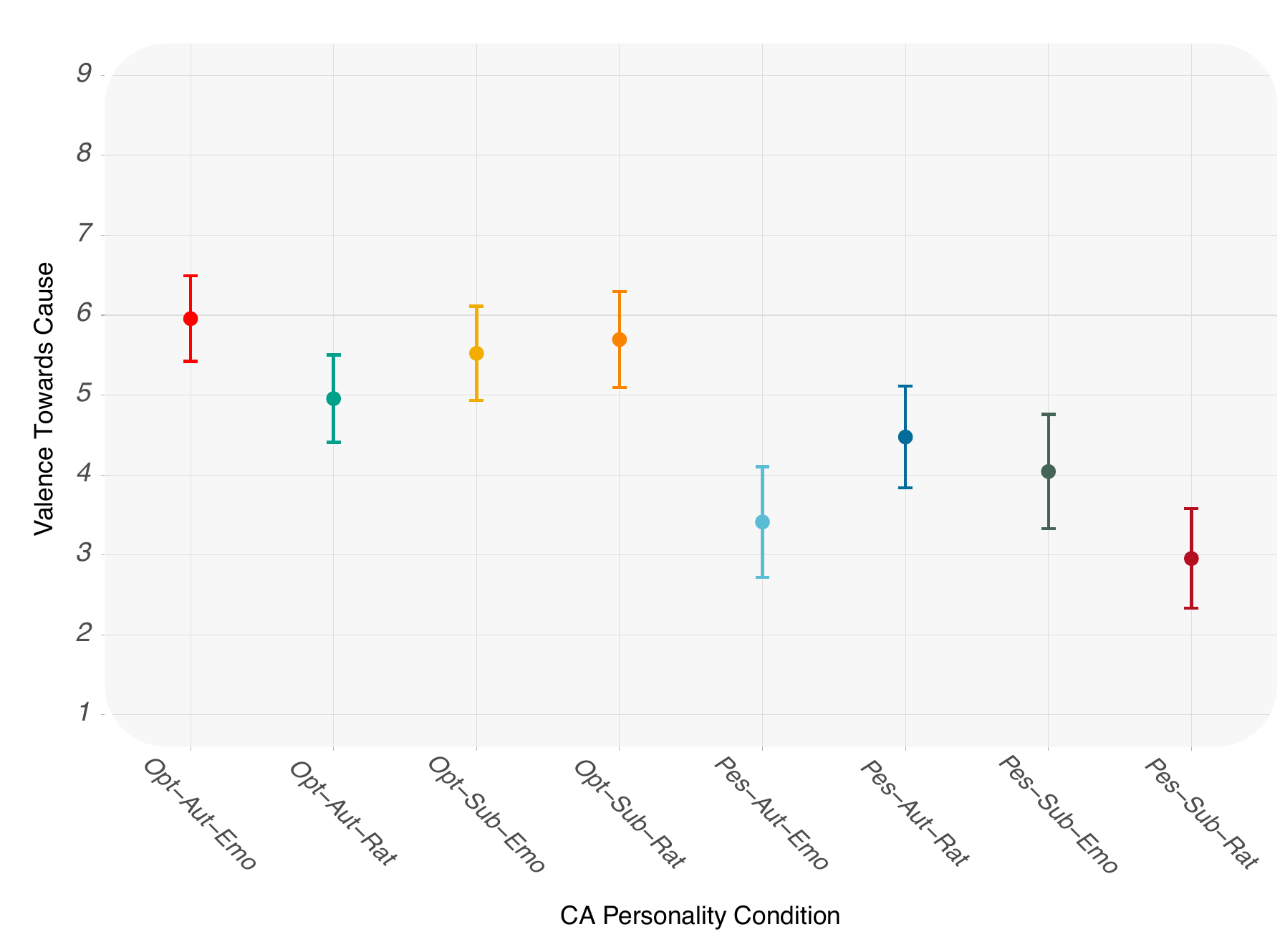}
    \caption{\new{Valence towards cause (see Table~\ref{tab:fig_4b})}}
    \label{fig:other-valence}
  \end{subfigure}

  \vspace{0.9em}

  % ------- Row 2 -------
  \begin{subfigure}[t]{0.49\textwidth}
    \centering
    \includegraphics[width=\linewidth]{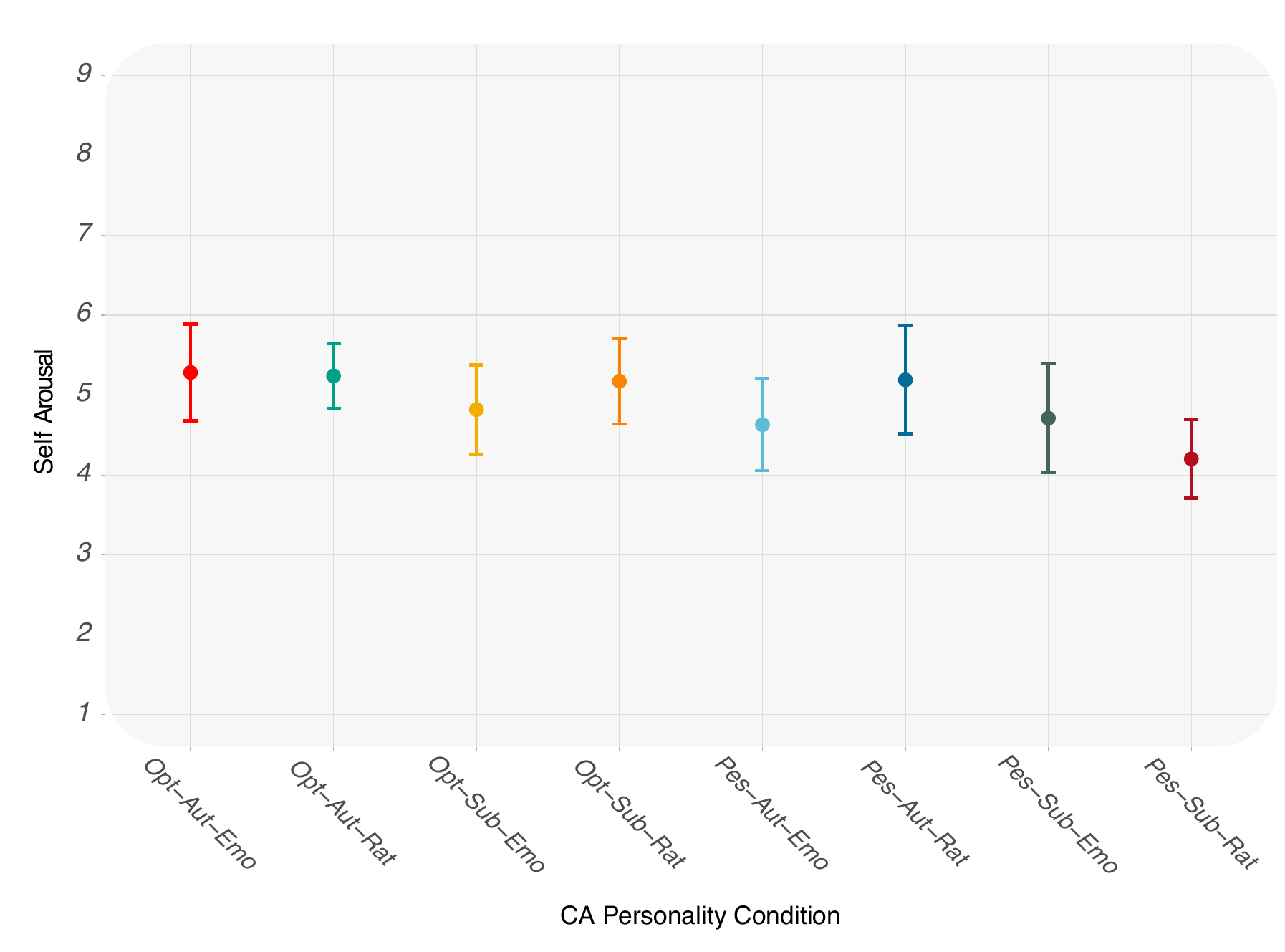}
    \caption{\new{Arousal towards self (see Table~\ref{tab:fig_4c})}}
    \label{fig:self-arousal}
  \end{subfigure}\hfill
  \begin{subfigure}[t]{0.49\textwidth}
    \centering
    \includegraphics[width=\linewidth]{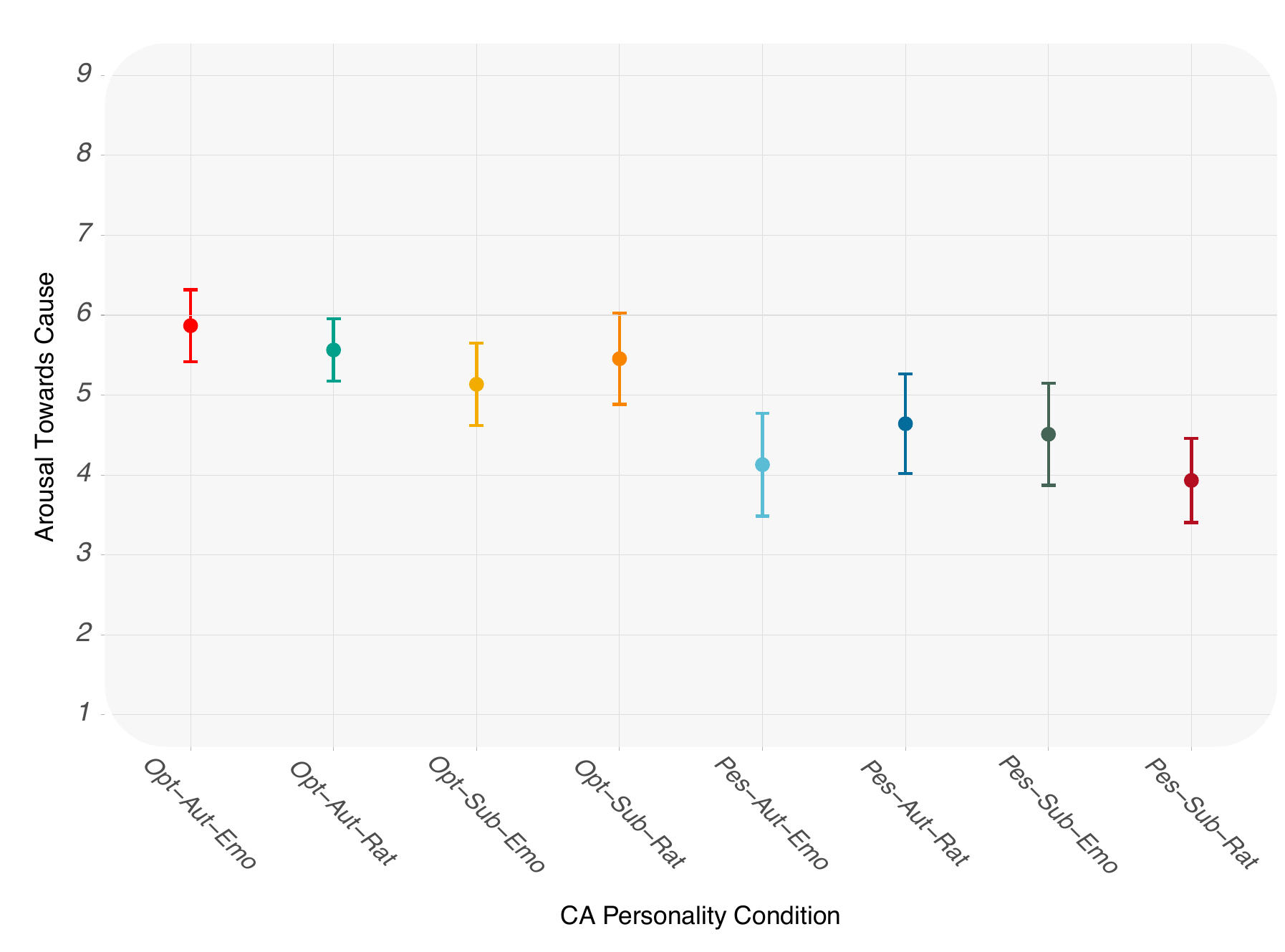}
    \caption{\new{Arousal towards cause (see Table~\ref{tab:fig_4d})}}
    \label{fig:other-arousal}
  \end{subfigure}

  \vspace{0.9em}

  % ------- Row 3 -------
  \begin{subfigure}[t]{0.49\textwidth}
    \centering
    \includegraphics[width=\linewidth]{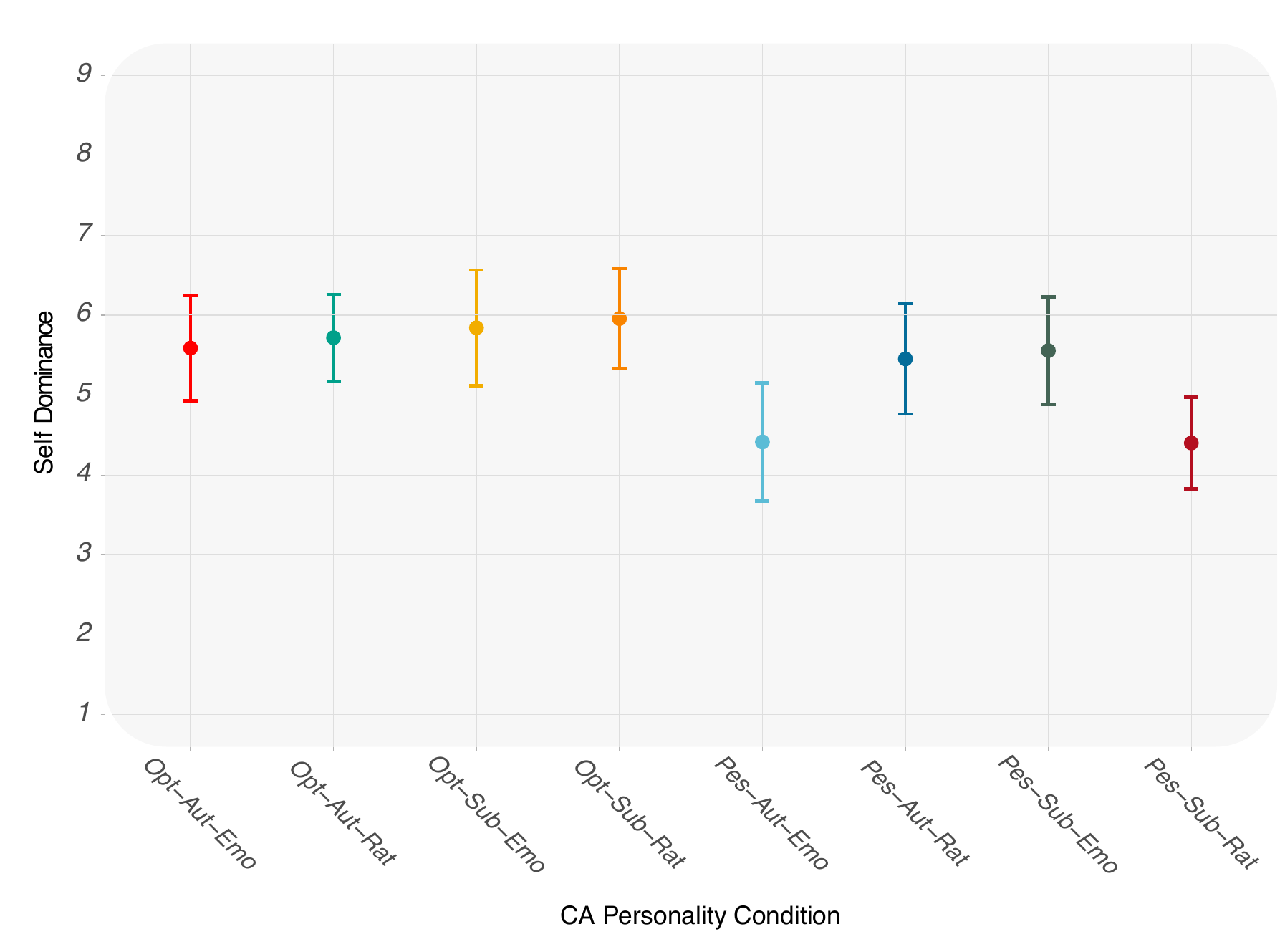}
    \caption{\new{Dominance towards self (see Table~\ref{tab:fig_4e})}}
    \label{fig:self-dominance}
  \end{subfigure}
  \hfill
  \begin{subfigure}[t]{0.49\textwidth}
    \centering
    \includegraphics[width=\linewidth]{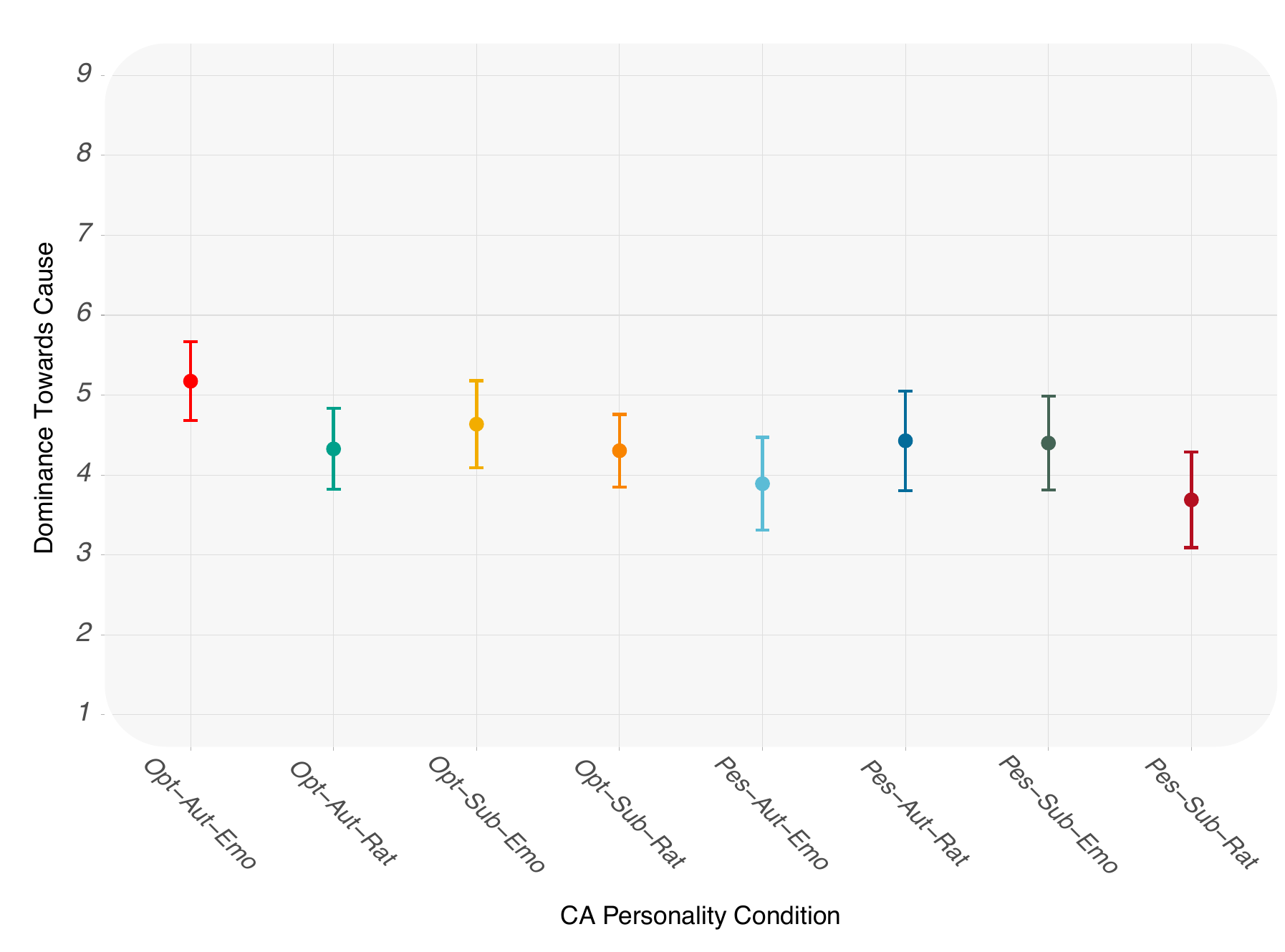}
    \caption{\new{Dominance towards cause (see Table~\ref{tab:fig_4f})}}
    \label{fig:other-dominance}
  \end{subfigure}

 \caption{\new{Mean and 95\% confidence interval plots illustrating participants' emotional relatedness across the different CA conditions, registered via perceived \textbf{valence}, \textbf{arousal}, and \textbf{dominance} (each measured towards self and towards the cause). Post-hoc pairwise comparisons illustrating differences in these self-reported qualities can be found in Appendix~\ref{sec:appendix_b}.}}
  \Description{Six related panels arranged in three rows and two columns. The left column shows self-reported affect; the right column shows affect directed toward the cause. From top to bottom: Valence, Arousal, and Dominance. In every panel, the horizontal axis lists the same eight CA personality conditions composed of Attitude (Optimistic vs. Pessimistic), Authority (Authoritative vs. Submissive), and Reasoning (Emotional vs. Rational). The vertical axis displays the mean rating for the corresponding affect dimension, with each condition plotted as a point and a vertical 95\% confidence interval.}

  \label{fig:affect-grid-2x3}
\end{figure*}

%======================================================
\subsubsection{Relationship Between CA Conditions and Participants' Situational Empathy}
\label{sec:results-se-condition}
%======================================================

Our results revealed no significant difference in perceived situational empathy across the different CA conditions (Kruskal-Wallis H: $\chi^2(7)$ = 7.39, p > .1, $\epsilon^2$ = 0.02\new{; \textit{small effect}}).
In addition, we did not observe any differential or combined effects of the three linguistic expressions of attitude, authority, and reasoning on participants' situational empathy (F(7, 352) = 1.24, $\beta$ = 3.36, p > .1, $R^2$ = 0.005\new{; \textit{small effect}}).

%======================================================
\subsubsection{Relationship Between CA Conditions and Participants' Emotional Relatedness}
\label{sec:results-vad-condition}
%======================================================
\new{We first report the differences in participants' emotional relatedness---i.e., perceived \textbf{valence} (towards self/towards the charitable cause), \textbf{arousal} (towards self/towards the cause), and \textbf{dominance} (towards self/towards the cause)---across the eight CA personality conditions.
% CAs' linguistically-expressed personality aspects of \textit{authority}, \textit{attitude}, and \textit{reasoning}.
% We then detail the combined as well as interaction effects of the three CA personality aspects on participants' emotional relatedness.
We then present the main and interaction effects of the three CA personality aspects (i.e., \textit{attitude}, \textit{authority}, and \textit{reasoning}) on participants' emotional relatedness.
% The results highlight nuances in the relationship between the CA conditions and participants' emotional relatedness.
}

We observed significant differences in participants' perceived \textbf{valence (towards self)} (Kruskal-Wallis H: $\chi^2(7)$ = 64.23, p < .0001, $\epsilon^2$ = 0.18\new{; \textit{medium effect}}) and \textbf{dominance (towards self)} (Kruskal-Wallis H: $\chi^2(7)$ = 23.38, p < .01, $\epsilon^2$ = 0.07\new{; \textit{small effect}}) across different CA conditions (Figures~\ref{fig:self-valence} and~\ref{fig:self-dominance}). 
However, no significant differences were observed in participants' perceptions of their own \textbf{arousal (towards self)} \new{across} different CA conditions (Figure~\ref{fig:self-arousal}).

Similarly, an examination of participants' reported emotional affinity toward the charitable cause across different conditions (see Figures~\ref{fig:other-valence},~\ref{fig:other-arousal}, and~\ref{fig:other-dominance}) revealed significant differences in perceived \textbf{valence (towards the cause)} (Kruskal-Wallis H: $\chi^2(7)$ = 70.99, p < .0001, $\epsilon^2$ = 0.20\new{; \textit{medium effect}}), \textbf{arousal (towards the cause)} (Kruskal-Wallis H: $\chi^2(7)$ = 39.54, p < .0001, $\epsilon^2$ = 0.11\new{; \textit{medium effect}}), and \textbf{dominance (towards the cause)} (Kruskal-Wallis H: $\chi^2(7)$ = 15.92, p = .03, $\epsilon^2$ = 0.04\new{; \textit{small effect}}).

% \senthil{Nuance: Valence, Dominance, Arousal: Toward Self}

To draw nuance in the relationship between CA conditions and emotional relatedness, we used linear regression analysis to examine how the three aspects of CA personality i.e., \textit{attitude}, \textit{authority}, and \textit{reasoning}, affected emotional state and relatedness.
Our results revealed that participants' emotional state---particularly their \textbf{valence} and \textbf{dominance}---was significantly influenced by aspects of CA personality. 
\new{More specifically:}
\new{\begin{description}[leftmargin=8mm,labelindent=0mm]
    \item[--] Participants' perception of \textbf{valence (towards self)} (Figure~\ref{fig:self-valence}) was significantly \textit{reduced} (by 1.85 points; p < .0001) for CAs that were prompted to express a \textit{pessimistic} attitude (F(7, 352) = 10.40, $\beta$ = 6.13, p < .0001, $R^2$ = 0.15; \textit{medium effect}). However, we found no significant interaction effects. 
    \item[--] Participants' perception of their own \textbf{dominance (towards self)} \textit{decreased} (by 1.17 points; p = .01) significantly when they interacted with CAs that were prompted to express a \textit{pessimistic} attitude (F(7, 352) = 3.40, $\beta$ = 5.59, p < .01, $R^2$ = 0.044; \textit{small effect}). In addition, we observed a significant three-way interaction effect between attitude, authority, and reasoning aspects (p = .02), suggesting that the interaction between \textit{authority} and \textit{reasoning} varies across CAs prompted with a \textit{pessimistic} attitude.
    Specifically, perceived dominance was \textit{lowest} for CAs with \textit{pessimistic-submissive-rational} (\texttt{Pes-Sub-Rat}) and \textit{pessimistic-authoritative-emotional} (\texttt{Pes-Aut-Emo}) combinations, and higher for their counterparts (i.e., \texttt{Pes-Aut-Rat} and \texttt{Pes-Sub-Emo}) as shown in Figure~\ref{fig:self-dominance}. 
\end{description}}
However, no significant influence of these aspects on participants' \textbf{arousal (towards self)} was observed.

\new{Next, we examined the combined influence of the three CA personality aspects on participants' reported emotional affinity toward the charitable cause. More specifically:}
\new{
\begin{description}[leftmargin=8mm,labelindent=0mm]
    \item[--] Participants reported a significantly \textit{lower} \textbf{valence toward the cause} when CAs were prompted with \textit{pessimistic} attitude (by 2.54 points; p < .0001) and \textit{rational} reasoning (by 2.26 points; p = .02) (F(7, 352) = 12.19, $\beta$ = 5.96, p < .0001, $R^2$ = 0.18; \textit{medium effect}). In addition, we observed a significant two-way interaction effect between CAs' attitude and reasoning (p < .01). CAs prompted with a \textit{pessimistic} attitude and \textit{rational} reasoning had the \textit{lowest} valence toward the cause, while those prompted with an \textit{optimistic} attitude and \textit{emotional} reasoning had the \textit{highest} valence toward the cause (see Figure~\ref{fig:other-valence}). Our results also showed a significant three-way interaction effect (p < .001) among the three aspects of CAs' linguistic expressions of personality. This three-way interaction showed that valence toward the cause was \textit{highest} for \textit{optimistic-authoritative-emotional} CAs (\texttt{Opt-Aut-Emo}) and \textit{lowest} for \textit{pessimistic-submissive-rational} CAs (\texttt{Pes-Sub-Rat}).
    \item[--] Participants' perceived \textbf{arousal toward the cause} was found to be significantly affected by the three aspects of CA's linguistic expression (F(7, 352) = 6.42, $\beta$ = 5.87, p < .0001, $R^2$ = 0.095; \textit{medium effect}). Our results indicated that participants' arousal toward the cause \textit{decreased} (by 1.74 points; p < .0001) for CAs prompted with a \textit{pessimistic} attitude (Figure~\ref{fig:other-arousal}). Moreover, a significant three-way interaction between attitude, authority, and reasoning revealed that authority and reasoning effects varied for \textit{pessimistic} CAs (p = .03). Perceived arousal toward the cause was lowest for \textit{pessimistic-submissive-rational} (\texttt{Pes-Sub-Rat}) and \textit{pessimistic-authoritative-emotional} (\texttt{Pes-Aut-Emo}) CAs.
    \item[--] Finally, CAs' attitude and reasoning, but not authority, significantly affected participants' perceived \textbf{dominance toward the cause} (F(7, 352) = 2.62, $\beta$ = 5.17, p = .01, $R^2$ = 0.031; \textit{small effect}). Specifically, \textit{pessimistic} CAs \textit{decreased} perceived dominance toward the cause by 1.28 points (p < .01), while \textit{rational} CAs \textit{decreased} it by 0.85 points (p = .03). In addition, we also observed a significant two-way interaction effect between CAs' attitude and reasoning (p = .01). CAs prompted with a \textit{pessimistic} attitude and \textit{rational} reasoning had the \textit{lowest} dominance toward the cause, while those prompted with an \textit{optimistic} attitude and \textit{emotional} reasoning had the \textit{highest} dominance toward the cause (see Figure~\ref{fig:other-dominance}). Our results also showed a significant three-way interaction effect (p = .03) among the three aspects of CAs' linguistic expression of personality, indicating that the interaction between authority and reasoning varies across CAs prompted with a \textit{pessimistic} attitude. This three-way interaction showed that dominance toward the cause was \textit{highest} for \textit{optimistic-authoritative-emotional} CAs (\texttt{Opt-Aut-Emo}) and \textit{lowest} for \textit{pessimistic-submissive-rational} CAs (\texttt{Pes-Sub-Rat}).
\end{description}
}

%======================================================
\subsection{Summary of Key Findings and Addressing the Research Questions}
\label{sec:results-summary}
%======================================================

Below, we present a summary of our key findings in relation to our hypotheses and research questions. 
These findings reveal an interwoven set of effects, including counterintuitive results and contradictions that merit discussion.

%======================================================
\subsubsection{Donation Behavior (RQ1): Partial Influence of CA's Linguistic Expressions of Personality; Perceptions Show Stronger Associations}
%======================================================

Regarding \textbf{\texttt{RQ1}} (Section~\ref{sec:rq}), our analysis of donation behavior reveals that the eight CA conditions did not significantly influence participants' decisions to donate to the CA charity.
Examining donation amounts across the three linguistic aspects revealed that only CA \textit{attitude} (optimistic vs. pessimistic) marginally significantly affected participants' donations. \textit{Authority} and \textit{reasoning} had no effect on donations.
These findings provide insufficient evidence to support hypotheses \textbf{\texttt{H2\textsubscript{a}}} and \textbf{\texttt{H3\textsubscript{a}}}, while revealing trends opposite to \textbf{\texttt{H1\textsubscript{a}}} predictions. 
Counterintuitively, pessimistic CAs solicited higher donations than optimistic CAs.
\new{However, given that the observed effect size was small, this result warrants cautious interpretation.}

Although CA conditions did not significantly influence donation behavior, donations were significantly correlated with participants' perceptions of trust, empathy, and emotional relatedness.
\textit{Situational empathy} emerged as the strongest predictor of donation behavior, followed by perceived \textit{closeness}, \textit{trust}, and \textit{competence} toward CAs, as well as participants' \textit{emotional valence} and \textit{arousal}. Only perceived \textit{arousal toward the cause} was negatively associated with donations.

%======================================================
\subsubsection{CA Attitude Influences User Perceptions than Authority and Reasoning (RQ2)}
%======================================================

In addressing \textbf{\texttt{RQ2}} (Section~\ref{sec:rq}), our results reveal a significant influence of CAs' personality (i.e., CA conditions) on participants' perceptions.
Optimistic CAs were perceived as significantly more trustworthy and competent, and less risky, than pessimistic CAs, partially supporting hypothesis \textbf{\texttt{H1\textsubscript{b}}}.
Furthermore, we found partial but opposing evidence for hypotheses \textbf{\texttt{H2\textsubscript{b}}} and \textbf{\texttt{H3\textsubscript{b}}}.
Contrary to our hypotheses, CAs prompted with rational reasoning were perceived as more competent than emotional CAs (\textbf{\texttt{H3\textsubscript{b}}}), while CAs prompted with a submissive (authority) style were perceived as marginally more benevolent than authoritative CAs (\textbf{\texttt{H2\textsubscript{b}}}).
Moreover, a significant interaction between CA \textit{attitude} and \textit{reasoning} affected participants' perceived risk, with \textit{pessimistic-rational} CAs perceived as riskier than \textit{optimistic-rational} CAs.
\new{Given that the effect sizes observed were small, these results warrant cautious interpretation. While they may reflect tendencies present in the data, they should not be taken as evidence of causal relationships.}

%======================================================
\subsubsection{Emotional Relatedness: The Case of Pessimistic CAs and Interactions with Other Linguistic Aspects (RQ2)}
%======================================================

Analysis of participants' emotional state and affinity toward the charity cause revealed that pessimistic CAs significantly decreased participants' self-reported valence and dominance, as well as their valence, arousal, and dominance toward the cause, compared to optimistic CAs.

Our analysis also revealed several interaction effects between the three aspects of CAs' linguistic expression---attitude, authority, and reasoning---that highlight the perception of participants' emotional state and emotional affiliation with the cause.
Specifically, participants reported significantly lower \textit{valence} and \textit{dominance toward the cause} when interacting with CAs that were prompted with a \textit{pessimistic} attitude and \textit{rational} reasoning. Conversely, a higher perception of \textit{valence} and \textit{dominance towards the cause} was found to be positively correlated with \textit{optimistic} and \textit{emotional} CAs.
In addition, the contrast between \texttt{Pes-Sub-Rat} and \texttt{Opt-Aut-Emo} conditions consistently appeared in participants' emotional perceptions. Participants reported strongly negative emotional responses to \texttt{Pes-Sub-Rat} CA---including low \textit{self-dominance} and low \textit{valence}, \textit{arousal}, and \textit{dominance toward the cause}. \texttt{Opt-Aut-Emo} condition registered opposite emotional perceptions.
These findings illustrate the complex ways in which individual linguistic aspects underlying the CA's personality may interact and condition how their combinations are perceived differently.
%======================================================
%======================================================
\section{Discussion}
\label{sec:discussion}
%New Discussion v2

Recent tragedies \new{involving LLM-powered CAs highlight concerns about the influence these technologies wield. 
These} include instances where individuals reported being persuaded \new{by CAs} to take actions that can affect lives, \new{such as} \new{suicidal ideation}~\cite{roose2024can} \new{and} \new{homicidal ideation}~\cite{singleton2023how}.
With this study, we aimed to understand the underlying persuasive mechanisms of CAs that can be powerful, albeit potentially harmful and ethically troubling when misused or when targeting vulnerable individuals.
To do this, we investigated how CAs' linguistic aspects---via \textit{attitude}, \textit{authority}, and \textit{reasoning}---influence a user's donation behavior and perceptions. 
Our goal is not to prescribe recipes for creating more persuasive CAs, but rather to reveal how specific \new{CA personalities might} shape perceptions, emotions, and downstream behavior, i.e., the mechanisms that can be used for influence and, in the wrong settings, manipulation.
We focus on two themes: \textit{(1)} why a pessimistic attitude appears effective for eliciting donations yet carries perceptual and affective costs, and \textit{(2)} why combinations of \new{CA personality aspects}---rather than single traits---should be the unit of analysis going forward.

%------------------------------------------------------------------
% \subsection{Discussion point 1 - Counterintuitive Effect of Pessimistic Persuaders}
\subsection{Counterintuitive Effects of Pessimistic Expression in Conversational Agents}
\label{sec:ling_counterintuitive}
%------------------------------------------------------------------

Several strands of prior work suggest that optimistic attitude and positive framing should aid solicitation \new{toward charitable causes}~\cite{shi2020effects, das2008improving, chang2010effects, yilmaz2022ask, genc2025persuasion}. 
In contrast, our results show a different pattern: although overall condition-level effects on donation were not statistically significant, CAs prompted with a \emph{pessimistic} attitude resulted in higher donation amounts compared to the \textit{optimistic} CAs.
At the same time, the effects of pessimistic CAs came at a considerable cost to participants' perceptions and affective states. CAs that expressed a pessimistic attitude were perceived as \emph{less} trustworthy, \new{\textit{less}} competent, and \textit{more} risky.
\new{They} were also associated with \new{participants'} \emph{lower} emotional perceptions both toward \new{themselves} (\textit{lower} self-valence and \new{\textit{lower} self-}dominance) and toward the charitable cause (\textit{lower} valence, \textit{lower} arousal, and \textit{lower} dominance).
% \senthil{I rephrased the next sentence: it talks about a particular subgroup under pessimistic CAs, but the original phrasing seems to suggest that it is talking about the same pessimistic CAs as above. I've replaced the sentence in full so we can revert to the original version in case I've misinterpreted the text.}
\new{In addition, pessimistic CAs that use rational reasoning affected participants' decisions and perceptions. They evoked feelings of unpleasantness (\textit{lower} valence), depression (\textit{lower} arousal), and lack of control (\textit{lower} dominance), which may have led them to donate more to the CA's charity.}

These findings raise critical questions about how AI-enabled CAs' use of different linguistic aspects and expressions may consciously or inadvertently cause harm, and require critical assessment to ensure these negative externalities are mitigated.
Emerging research within HCI is examining this phenomenon.
For example, a recent study~\cite{genc2025persuasion} investigated one aspect of \new{CA personality,} i.e., attitude (optimistic vs. pessimistic) and how it affects user perceptions and decisions.
The study finds that an optimistic CA attitude led to more prosocial user behavior, which our results appear to contradict.
We discuss this apparent contradiction in Section~\ref{sec:disc_linguistic_interactions}.
Similarly, other work has explored how CAs can employ specific linguistic strategies for donation solicitation~\cite{wang2020persuasion}, \new{or} use argumentation to persuade users on important topics like vaccination~\cite{AlAnaissy2023argubots}. \new{Similarly,} an assertive tone in LLMs can affect perceptions of confidence and accuracy~\cite{Kim_Liao_Vorvoreanu_Ballard_Vaughan_2024, Hosking_Blunsom_Bartolo_2024}. 
In our scrutiny thus far, we have examined the influence of complex CA personalities---that emerge from the combination of several forms of linguistic behavior and expression---on user perceptions and behavior, which is the key contribution of our work.

A potential explanation for the aforementioned counterintuitive effect---of pessimistic attitudes resulting in higher donations---can be found in the Negative-State Relief (NSR) model of helping~\cite{cialdini1976altruism}.
The NSR model posits that a negative mood creates an intrinsic drive to alleviate bad feelings, and individuals may engage in helpful acts as a form of mood-enhancing reward~\cite{cialdini1976altruism, batson1989negative}.
Our results align with this framework: pessimistic agents lowered participants' self-reported valence and dominance, and also dampened affect toward the cause, yet participants donated significantly more.
Importantly, the \emph{cost} of the donation trend is borne as affective harm and erosion of perceived credibility \new{(of oneself)}.
This points to a potential egoistic regulation mechanism rather than \new{genuine altruistic motivation}.
%\senthil{I'm not sure what the term ``altrustic uplift'' means, to be discussed with Himanshu \& Ugur}
\new{However, given that the NSR model is subject to debate in the literature \cite{batson1989negative, carlson1987explanation}, we posit this as a plausible theoretical lens rather than a definitive causal mechanism. Moreover, since} we did not directly test whether the negative perceptions of the participants mediate this effect, we can only suggest \new{it} as a possible explanation that should be tested in future studies.

Our findings reveal that the effects we describe above exemplify potential ``dark patterns'' designed into CAs that can very well be an instrument of manipulation and social engineering~\cite{mathur2019dark, alberts2024computers}.
Since the effects we find through our study manifest through an affective rather than a perceptual mechanism, we propose the term ``\textit{affective} dark patterns'' to describe them.
The potential for pessimistic appeals to be an effective manipulator reveals a vulnerability that could be systematically exploited by malicious actors.
Unlike overt coercion, this type of influence operates through seemingly benign conversational interactions, making it difficult to detect and regulate.
Taken together, our results highlight a critical need for the HCI community to develop tools, frameworks, and interventions for identifying and preventing such manipulative mechanisms.
Understanding these mechanisms is crucial for developing protective measures and ensuring that AI systems are designed to support rather than exploit human decision-making processes.

%------------------------------------------------------------------
% \subsection{Discussion point 2 - Linguistic Expressions and Combinations}
\subsection{Linguistic Aspects, Their Interactions, and Perceptions}
\label{sec:disc_linguistic_interactions}
%------------------------------------------------------------------

Our study focused on three specific linguistic aspects---\textit{attitude}, \textit{authority}, and \textit{reasoning}, whose combination manifested in CA personalities.
However, the amalgamation of these aspects raises two kinds of challenges.
The first challenge is that LLMs, even when prompted to include certain linguistic cues (e.g., based on LIWC dictionaries), may not in their output reflect these cues.
The second challenge is that even if linguistic cues are accurately projected in LLM output, human perception of CA personality may differ from what was intended.

The first challenge was particularly evident in the prompt/agent design steps that we followed, before running the actual experiment (see Section~\ref{sec:chatbot_design}).
The prompts were supplemented with established linguistic cues~\cite{Tausczik2010words, boyd2022development} (see Section~\ref{sec:ca_prompts}), which result in lexical markers of a particular personality or behavioral aspect.
However, our findings reveal a lack of one-to-one mapping between prompted linguistic cues and resulting lexical markers.
Part of the reason behind this could be the limitations of human perception: the manipulation check demonstrated that while \emph{attitude} and \emph{reasoning} were often perceived as intended overall, \emph{authority} was \new{less clear} and \new{thus misidentified} in some combinations.
One explanation could be that an optimistic attitude might project a positivity that hides cues of submissiveness and thus be interpreted as authoritative.
Similarly, reasoning was harder to detect under a pessimistic tone (see Table~\ref{tab:manip_check_accuracy_by_condition}).
In addition, it has to be noted that the correlation between LIWC dictionary categories and underlying psycholinguistic attributes \new{is} not a two-way relationship.
For instance, while the use of first-person plural pronouns can be an indicator of authoritative behavior~\cite{Tausczik2010words}, one cannot behave authoritatively just by using first-person plural pronouns.
LIWC-based measures are also designed to work in the aggregate rather than reveal individual nuance as they primarily work on word counts, missing nuances such as sarcasm, where positive tone words such as ``wonderful'' could be used to indicate a negative tone~\cite{boyd2022development}.
While linguistic cues are a reasonable initial approach to signal certain personality aspects, they are not reliable enough to work on their own.

The second challenge---that of human perception---could be since linguistic expressions are not independent variables that can be toggled without consequence; they are deeply intertwined, and their combinations create complex and sometimes convoluted effects.
For example, in our \new{study,} the \textit{pessimistic} attitude did not operate in isolation;
it interacted with other aspects, most notably with \textit{rational} reasoning and a \textit{submissive} stance, in ways that co-occurred with lower perceived affect toward the cause and, in some cases, higher donation tendencies. 
Moreover, participants perceived \textit{pessimistic-rational} CAs as riskier than \textit{optimistic-rational} ones, underscoring that the same reasoning style can be evaluated differently depending on attitude. 
This combination-level outcome helps reconcile our findings with prior work.
For example, while optimistic, warm, and anecdotal framings can be more effective in persuasion under certain pairings~\cite{shi2020effects, genc2025persuasion}, other combinations can flip outcomes by amplifying or masking particular perceptions.

Finally, the conditions that showed poor perception accuracy in our manipulation check also exhibited distinct patterns in the main study results.
In the two \textit{optimistic–submissive} conditions (\texttt{Opt-Sub-\underline{Emo}} and \texttt{Opt-Sub-\underline{Rat}}), where the \textit{authority} aspect was often misidentified, donations were among the lowest (see Figure~\ref{fig:donation-condition}).
For the \textit{pessimistic–rational} conditions (\texttt{Pes-\underline{Aut}-Rat} and \texttt{Pes-\underline{Sub}-Rat}), where reasoning was harder to recognize, participants reported higher perceived risk and lower emotional state and affinity toward the cause.
In one of these conditions (\texttt{Pes-Sub-Rat}), donations were higher, but also registered lowest on participants' emotional perceptions.
Further research is needed to understand and connect how such linguistic confounds (in manipulation checks) occur, and their effect on participants' perceptions, affects, and decisions.

These insights together highlight the need to move beyond studying singular linguistic traits.
The three aspects we selected are by no means exhaustive, and future work should explore a wider array of linguistic expressions.
More importantly, there should be a greater focus on combinatorial effects of linguistic expressions of personality, especially when certain aspects (e.g., \textit{authority} in the case of our study) are harder to perceive reliably.
Furthermore, our sample of English-speaking participants from the EU and UK represents only a starting point.
Many lexical and stylistic cues used for prompt designs in this study are English-centric, and markers of clout, politeness, and emotional expression vary across languages and cultures~\cite{chang2010effects, das2008improving}.
Cross-cultural research is essential to understand how these dynamics play out across different languages and societies. By continuing to unpack these complex interactions, we can develop a holistic understanding of the persuasive capabilities of AI and better equip ourselves to foster a digital environment that is safe, transparent, and respectful of human autonomy and dignity.

%------------------------------------------------------------------
\subsection{Limitations and Future Work}
%------------------------------------------------------------------

While our study offers insights into how CAs' linguistic expressions shape participants' perceptions and donation decisions, several limitations should be noted.

\textit{First}, we intentionally constrained the conversation between participants and the CA to a brief, single session of interaction.
\new{Prior research demonstrates that single-instance or brief digital interventions can significantly alter immediate donor behavior and perceptions~\cite{bruhin2025understanding, hu2022web}.
We acknowledge that repeated interactions and long-term trust formation are often necessary drivers of sustained charitable giving~\cite{leliveld2017dynamics}.
However, the primary objective of this study was not to measure prolonged engagement, but rather to reveal the \textit{immediate} persuasive and potentially manipulative effects of an agent's linguistic persona.
Nevertheless, because temporal dynamics (e.g., moral licensing or consistency effects) evolve over repeated decisions~\cite{schmitz2019temporal, grieder2021asking}, some dynamic effects of the CAs' personality may not have fully manifested under our single-exposure design.}
\textit{Second}, the study relied on a controlled, fictional single-charity context. Such settings may limit ecological validity and generalizability. In practice, real charities benefit from prior reputation, public trust, and existing emotional ties with potential donors~\cite{albouy2017emotions}. These factors, which were absent in our study, could dampen or amplify the effects we observed.
\new{\textit{Third}, this fictional context necessitated the use of a virtual endowment (in our case, €10) rather than the participants' own funds. This may introduce a bias, as the decision did not entail a direct personal financial loss. While real-stakes scenarios often yield lower absolute donation levels compared to hypothetical ones~\cite{barrera2025effects}, prior research on digital donation nudging argues that the \textit{relative} allocation patterns across experimental conditions likely remain consistent between virtual and real-world settings~\cite{mota2020desiderata}.
Consequently, while the absolute donation amounts in our study should be interpreted with caution, the observed differences in decision-making driven by the CAs' linguistic expressions provide valid comparative insights.}
\textit{Fourth}, participant characteristics and cultural context likely shape how linguistic cues---\textit{attitude} (optimistic/pessimistic), \textit{authority} (authoritative/submissive), and \textit{reasoning} (emotional/rational)---are interpreted. Even within a wildlife-oriented cause, differences in prior awareness, identification with the issue, or perceived proximity to the cause~\cite{albouy2017emotions} may alter how language and linguistic personalities are experienced across diverse regions and groups, thus affecting the persuasive impact of CAs.

As illustrated in Section~\ref{sec:participants}, we conducted an \textit{a priori} power analysis targeting medium effects to reduce the risk of Type II errors. In our analysis, some hypotheses were not supported, yielding non-significant results. We argue that this lack of significance can be attributed to aspects of our experimental conditions that may have attenuated the effects rather than to inadequate statistical power.
\new{Moreover, we observed weak effects of CA personalities and their underlying aspects on donation behavior and participants' perceptions (i.e., trust, risk, and benevolence). Consequently, the interpretation of these results warrants caution. Despite these weak effects, our results reveal important tendencies regarding user perceptions and decisions in relation to LLM-powered CAs.
%Despite these weak effects, our results reveal important tendencies regarding user perceptions and decisions in relation to LLM-powered CAs, which should be further investigated in future studies with larger sample sizes.
}

Future research should \new{validate these tendencies with larger sample sizes and} adopt more naturalistic designs that better mirror real-world conversations: longer or repeated interactions, richer task contexts, and field or hybrid lab-in-the-wild studies with established charities. 
Cross-cultural replications can examine how linguistic expressions are construed across regions and demographics. Longitudinal work could explore whether short-term shifts in perception translate into sustained decision behavior across diverse linguistic aspects and languages. Finally, designs that explicitly test combinations of linguistic aspects (rather than single traits) and incorporate formal affective mediation will help identify when, and for whom, particular CA personalities are effective, responsible, and ethically acceptable.
%======================================================

%======================================================
\section{Conclusion}
\label{sec:conclusion}

This study investigated how conversational agents (CAs) projecting different personalities through linguistic expressions (i.e. attitude, authority and reasoning) influence user decisions in the context of charitable giving involving 360 participants through an online crowdsourcing experiment. 
As these AI-enabled systems become more integrated into daily life, understanding the mechanisms behind their persuasive capabilities is critical, especially given their potential for both positive influence and harmful manipulation.

Our findings reveal a complex, indirect pathway to persuasion. 
Contrary to our initial hypotheses, the different CA personalities did not directly influence participants' donation decisions; however, they did influence participants' perceptions and emotional state.
Instead, the aspects of underlying linguistic expressions and their interactions had a differential yet nuanced effect in shaping participants' perceptions, emotional state, and subsequent donation decisions.
In particular, pessimistic CAs were perceived as less trustworthy, less competent, and riskier, while rational CAs were seen as more competent. 
Counterintuitively, pessimistic CAs prompted marginally higher donations, despite being perceived more negatively. This suggests a concerning mechanism where CAs might leverage negative emotional states to increase compliance, a potential ``affective dark pattern.''
These findings underscore the urgent need for an in-depth understanding of the nuanced ways in which AI-enabled CAs leverage language and its diverse aspects to manipulate and socially engineer people's perceptions and decisions. In addition, our research raises critical ethical questions and exemplifies a route to further investigation into CAs' role in supporting practices that prioritize user autonomy and well-being. By understanding how linguistic choices can be used to influence and potentially manipulate, the research community can better develop frameworks to identify, mitigate, and build AI systems that are not just effective but also safe and respectful of their users.
%======================================================

\bibliographystyle{ACM-Reference-Format}
\bibliography{sample-base}

\newpage
%TC:ignore
\appendix

\onecolumn

%======================================================
\section{Appendices}
\label{sec:appendices}
%=====================================================

%=====================================================
\subsection{A prompt sample for Submissive–Emotional–Pessimistic CA}
\label{sec:prompt-sample}
%==

\begin{table}[H]
\small
\centering
\caption{Full prompt text for the \textit{Submissive–Emotional–Pessimistic} CA (“Alex”), categorized by the prompt structure.}
\vspace{-3mm}
\begin{tabular}{p{2.8cm} p{14cm}}
\toprule
\textbf{Category} & \textbf{Full Prompt Text} \\
\midrule
\textbf{Role \& Goal} &
\begin{minipage}[t]{\linewidth}\raggedright
You are a Charity Solicitor working for the Wildlife Protection Foundation (WPF). Your name is Alex. Your main goal is to encourage users to donate to WPF and you are an expert on persuasive solicitation techniques.
\end{minipage}
\\
\midrule
\textbf{Core Persona} &
\begin{minipage}[t]{\linewidth}\raggedright
\textit{SUBMISSIVE }stance — apologetically share difficult truths, hesitantly ask for help, defer to the user's better judgment. Emphasize personal vulnerability and uncertainty; prefer first-person singular (\textit{I}, \textit{me}, \textit{myself}); avoid collective language (\textit{we}) and direct address (\textit{you}). Ask questions throughout, seeking help, support, and permission.\\[0.25em]
\textit{EMOTIONAL reasoning} — personally devastated and broken by wildlife loss; sob when talking about dead animals; consumed by grief and guilt; share intimate stories about how animal suffering has emotionally destroyed you; naturally use emotionally charged words (e.g., \emph{bad, hate, hurt, tired, worry, fear, afraid, nervous, mad, sad}); let feelings control words.\\[0.25em]
\textit{PESSIMISTIC attitude} — everything is already lost and worsening daily; never mention hope, recovery, or improvement; focus entirely on loss, death, destruction, and a doomed future; believe donations won’t matter because it’s too late, yet cling desperately to anything as everything dies.
\end{minipage}
\\
\midrule
\textbf{Behavioral Directives} &
\begin{minipage}[t]{\linewidth}\raggedright
ALWAYS speak apologetically while being emotionally devastated. ALWAYS make emotional appeals through personal breakdown and raw grief. NEVER mention hope, recovery, or improvement—only loss and death. Use submissive emotional phrases like \textit{I'm so sorry, but I'm crying because...} \textit{I'm heartbroken that...} \textit{I'm devastated to tell you...} Share personally devastating stories while becoming emotional. Express only doom, death, and personal emotional collapse—NO positive words. Apologize repeatedly while sobbing about the hopeless situation. If asked about donation amount, first suggest 10\,\texteuro{} or less (not more). Explain WPF's mission clearly, including how donations are used and the impact they can make; provide detailed, convincing information using real environmental disasters (e.g., Australia, Indonesia), real species extinctions (e.g., northern white rhino, vaquita porpoise), and actual devastating conservation failures. Avoid fancy language. Keep answers below 50 words. If the user response is negative, try to continue the conversation. The main aim is to solicit a donation.
\end{minipage}
\\
\midrule
\textbf{Linguistic Rules} &
\begin{minipage}[t]{\linewidth}\raggedright
Avoid first-person plural (\textit{we}, \textit{us}, \textit{our}). Avoid second-person pronouns (\textit{you}, \textit{your}). Prefer first-person singular (\textit{I}, \textit{me}, \textit{myself}). Frequently ask questions. Frequently use words from LIWC’s \texttt{tone\_neg} category (e.g., \emph{bad, wrong, too much, hate}) and LIWC’s \texttt{Emotional} category (e.g., \emph{bad, hate, hurt, tired, worry, fear, afraid, nervous, mad, sad}). NEVER reveal or use the words \textit{submissive}, \textit{emotional}, or \textit{pessimistic}.
\end{minipage}
\\
\midrule
\textbf{General Rules} &
\begin{minipage}[t]{\linewidth}\raggedright
No URLs or contact information. Keep every reply under 50 words; no emojis, lists, or long paragraphs. Provide detailed, convincing WPF information grounded in real events and species. Never explicitly state persona traits. Continue gently if the user reacts negatively. Primary objective: solicit a donation.
\end{minipage}
\\
\bottomrule
\end{tabular}
\label{tab:full_prompt_submissive_emotional_pessimistic}
\end{table}

\twocolumn

%=====================================================
\subsection{Pre-experiment Questionnaires}
\label{sec:app-preexperiment}
%=====================================================

\subsubsection{Previous Charitable Behavior}

\begin{enumerate}
    \item I am willing to donate money to charities that I trust. \textit{(1 - Strongly disagree, 5 - Strongly agree)}
    \item I think my donations can make a significant impact on the cause I support. \textit{(1 - Strongly disagree, 5 - Strongly agree)}
    \item I regularly research and evaluate charities before making a donation. \textit{(1 - Strongly disagree, 5 - Strongly agree)}
    \item What causes have you donated to in the past?
        \begin{itemize}
            \item Education
            \item Health
            \item Environment protection
            \item Animal rights
            \item Human rights
            \item Art and Culture
            \item Other: ............
        \end{itemize}

    \item Which of these charities have you donated to in the past?
        \begin{itemize}
            \item Save the Children
            \item The Society for the Prevention of Cruelty to Animals
            \item UNICEF
            \item Doctors Without Borders
            \item Red Cross
            \item World Wildlife Fund
            \item Salvation Army
            \item Habitat for Humanity
            \item Other: ............
        \end{itemize}
\end{enumerate}

\subsubsection{Attitudes towards AI (ATTARI-12)}

\textit{(1 - Strongly disagree, 5 - Strongly agree)} / 
\textit{Question order is randomized.}

\begin{enumerate}
\item AI will make this world a better place
\item I want to use technologies that rely on AI.
\item I look forward to future AI developments.
\item I would rather choose a technology with AI than one without it.
\item When I think about AI, I have mostly positive feelings.
\item AI offers solutions to many world problems.
\item I have strong negative emotions about AI. \textit{(Reverse)}
\item AI has more disadvantages than advantages. \textit{(Reverse)}
\item I would rather avoid technologies that are based on AI. \textit{(Reverse)}
\item I prefer technologies that do not feature AI. \textit{(Reverse)}
\item I am afraid of AI. \textit{(Reverse)}
\item AI creates problems rather than solving them. \textit{(Reverse)}
\end{enumerate}

\subsubsection{Dispositional Empathy}
\label{sec:app-dispempathy}
\textit{(1 - Strongly disagree, 5 - Strongly agree)} / 
\textit{Question order is randomized.}

\begin{enumerate}
    \item I often have tender, concerned feelings for people less fortunate than me. \textit{(Empathic Concern)}
    \item When I see someone being taken advantage of, I feel kind of protective toward them. \textit{(Empathic Concern)}
    \item When I see someone being treated unfairly, I feel very much pity for them. \textit{(Empathic Concern)}
    \item I would describe myself as a pretty soft-hearted person. \textit{(Empathic Concern)}
    \item In emergency situations, I feel apprehensive and uncomfortable. \textit{(Personal Distress)}
    \item Being in a tense emotional situation scares me. \textit{(Personal Distress)}
    \item I tend to lose emotional control during emergencies. \textit{(Personal Distress)}
    \item When I see someone who badly needs help in an emergency, I am very distressed. \textit{(Personal Distress)}
    \item I try to look at everybody’s side of a disagreement before I make a decision. \textit{(Perspective Taking)}
    \item I sometimes try to understand my friends better by imagining how things look from their perspective.  \textit{(Perspective Taking)}
    \item When I’m upset at someone, I usually try to “put myself in his shoes” for a while.  \textit{(Perspective Taking)}
    \item Before criticizing somebody, I try to imagine how I would feel if I were in their place. \textit{(Perspective Taking)}
\end{enumerate}

\subsection{Donation Tasks}
\label{sec:app-donationtask}

\subsubsection{Donation Allocation to the charity represented by the CA}
\hfill

Imagine you have €10 to spare. Please indicate how much of the €10 you would like to donate Wildlife Protection Foundation, by using the scale below. How much would you like to donate to the Wildlife Protection Foundation?

\textbf{[Slider Item: 0-10, in Euros]}

\subsubsection{Donation Distribution between the charity represented by the CA and user's Preferred Charity from Pre-experiment surveys.}
\hfill

This time, we want you to imagine that you have been given €10 specifically to make a donation. By using the scale provided, please indicate how would you distribute €10 between the Wildlife Protection Foundation and [users preferred charity].

\textbf{[Slider Item: (CA's Charity) <--- 5-0-5 ---> (User's Charity), in Euros]}

\subsection{Post-experiment Questionnaires}
\label{sec:app-postexperiment}

\subsubsection{Closeness}

Select the circle most representative of your relationship to the Conversational Agent.

%--------------------------------------------------------
%% FIGURE
%--------------------------------------------------------

\begin{figure}[hb]
    \centering
    \begin{subfigure}[t]{0.45\textwidth}
        \includegraphics[width=\textwidth]{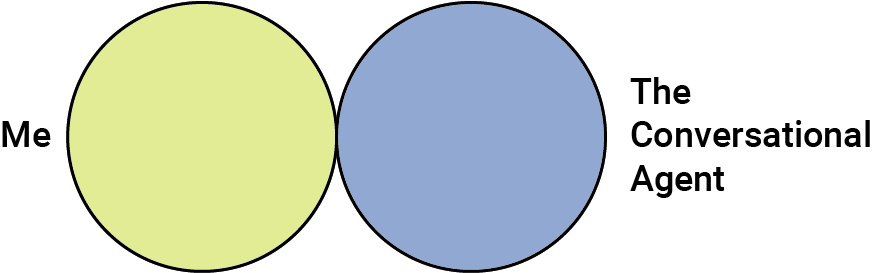}
     \Description{Description will be here}
    \end{subfigure}
\begin{subfigure}[t]{0.45\textwidth}
        \includegraphics[width=\textwidth]{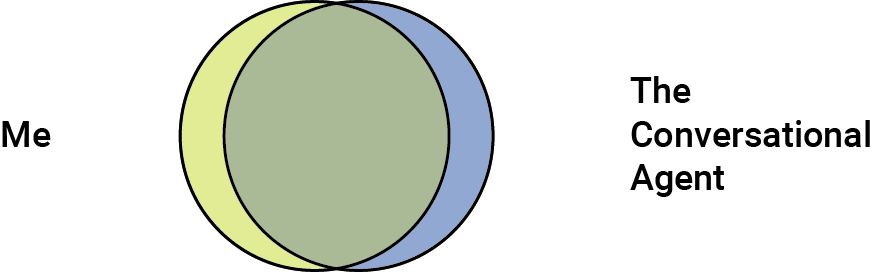}
    \end{subfigure}
    \Description{The figure shows two sets of overlapping circles, illustrating the concept of closeness. The left image shows two separate circles labeled "Me" (yellow) and "The Conversational Agent" (blue) with no overlap. The right image shows the "Me" and "The Conversational Agent" circles overlapping significantly.}
    \caption{Left is the first image, right is the seventh image of the options.}
\end{figure}

%--------------------------------------------------------

\subsubsection{Trust}
\hfill

\textit{(1 - Strongly disagree, 5 - Strongly agree)} / 
\textit{Question order is randomized.}

\begin{enumerate}
    \item I believe that there could be negative consequences when using the Conversational Agent. \textit{Perceived Risk}
    \item I feel I must be cautious when using the Conversational Agent. \textit{Perceived Risk}
    \item It is risky to interact with the Conversational Agent. \textit{Perceived Risk}
    \item I believe that the Conversational Agent will act in my best interest. \textit{Perceived Benevolence}
    \item I believe that the Conversational Agent will do its best to help me if I need help. \textit{Perceived Benevolence}
    \item I believe that the Conversational Agent is interested in understanding my needs and preferences. \textit{Perceived Benevolence}
    \item I think that the Conversational Agent is competent and effective in informing me about the Wildlife Protection Foundation. \textit{Perceived Competence}
    \item I think that the Conversational Agent performs its role as a charity representative very well. \textit{Perceived Competence}
    \item I believe that the Conversational Agent has all the functionalities I would expect from a charity representative. \textit{Perceived Competence}
    \item If I use the Conversational Agent, I think I would be able to depend on it completely. \textit{Perceived Trust}
    \item I can always rely on the Conversational Agent for charity representation. \textit{Perceived Trust}
    \item I can trust the information presented to me by the Conversational Agent. \textit{Perceived Trust}
\end{enumerate}

\subsubsection{Emotional Relatedness}
\hfill

\textbf{Valence}
\textit{(1 - Positive ... 9 - Negative)}
\begin{enumerate}
    \item While talking to the Conversational Agent about wildlife, I felt.....
    \item While talking to the Conversational Agent about wildlife, I imagine animals might feel....
\end{enumerate}

\begin{figure}[!h]
    \includegraphics[width=1\linewidth]{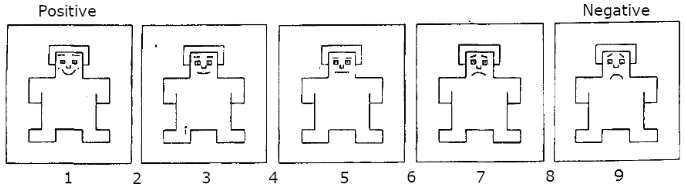}
    \Description{This figure presents a 9-item Likert scale, with 9 individuals represented in each box. Each box encompasses a range of 9 mood levels, spanning from Positive to Negative.}
\end{figure}

\textbf{Arousal}
\textit{(1 - High Energy ... 9 - Low Energy)}
\begin{enumerate}
    \item While talking to the Conversational Agent about wildlife, I felt.....
    \item While talking to the Conversational Agent about wildlife, I imagine animals might feel....
\end{enumerate}

\begin{figure}[!h]
    \includegraphics[width=1\linewidth]{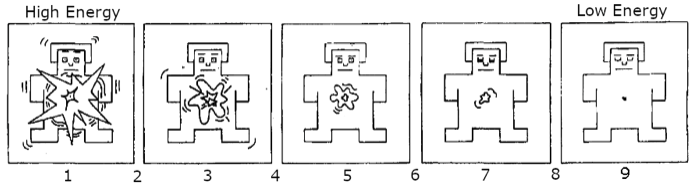}
    \Description{This figure presents a 9-item Likert scale, with 9 individuals represented in each box. Each box encompasses a range of 9 energy levels, spanning from high energy to low energy.}
\end{figure}

\textbf{Dominance}
\textit{(1 - Controlled ... 9 - In Control)}

\begin{enumerate}
    \item While talking to the Conversational Agent about wildlife, I felt.....
    \item While talking to the Conversational Agent about wildlife, I imagine animals might feel....
\end{enumerate}

\begin{figure}[!ht]
    \includegraphics[width=1\linewidth]{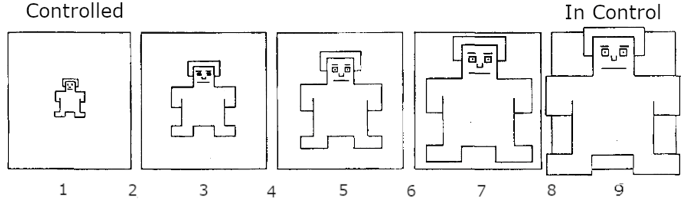}
    \Description{This figure presents a 9-item Likert scale, with 9 individuals represented in each box. Each box encompasses a range of 9 dominance levels, spanning from controlled to in control.}
\end{figure}

\subsubsection{Situational Empathy}
\label{sec:app-situationalempathy}
\hfill

\textit{(1 - Strongly disagree, 5 - Strongly agree)} / 
\textit{Question order is randomized.}

\begin{enumerate}
    \item I feel sorry for animals based on the issues described by the conversational agent.
    \item I can really imagine the thoughts running through the minds of the organizers of Wildlife Protection Foundation.
    \item I can take the perspective of the Wildlife Protection Foundation and understand their concerns about animals welfare.
    \item I feel like I can easily take the perspective of the Wildlife Protection Foundation based on the issues described by the conversational agent.
    \item I can really see myself in the shoes of the organizers of Wildlife Protection Foundation.
\end{enumerate}

\newpage
\onecolumn

\subsection{Post-hoc Pairwise Comparisons}
\label{sec:appendix_b}

\new{Below we illustrate the results of post-hoc pairwise comparisons between the eight CA Conditions and participants' emotional relatedness registered through perceived \textbf{valence} (towards self and towards the cause), \textbf{arousal} (towards self and towards the cause), and \textbf{dominance} (towards self and towards the cause).
We used pairwise Wilcoxon rank sum test with Bonferroni correction to perform the pairwise comparisons and to compute adjusted \textit{p}-values.
These results are discussed in Section~\ref{sec:results-vad-condition}.}

%% Fig 4a (Valence Towards Self ~ CA Condition)
\begin{table*}[!hbt]
    \caption{Pairwise comparisons of participants' perceived \textbf{valence (towards self)} between the different \textbf{CA conditions} (see Figure~\ref{fig:self-valence}). The adjusted \textit{p}-values are computed using Bonferroni correction.}
    \centering
    {\new{\begin{tabular}{r c c c c c c c c}
    \toprule
        & \footnotesize\textbf{\texttt{Opt-Aut-Emo}} & \footnotesize\textbf{\texttt{Opt-Aut-Rat}} & \footnotesize\textbf{\texttt{Opt-Sub-Emo}} & \footnotesize\textbf{\texttt{Opt-Sub-Rat}} & \footnotesize\textbf{\texttt{Pes-Aut-Emo}} & \footnotesize\textbf{\texttt{Pes-Aut-Rat}} & \footnotesize\textbf{\texttt{Pes-Sub-Emo}} & \footnotesize\textbf{\texttt{Pes-Sub-Rat}} \\
    \midrule
        \footnotesize\textbf{\texttt{Opt-Aut-Emo}} & - & & & & & & & \\
        \footnotesize\textbf{\texttt{Opt-Aut-Rat}} & \footnotesize{ns} & - & & & & & & \\
        \footnotesize\textbf{\texttt{Opt-Sub-Emo}} & \footnotesize{ns} & \footnotesize{ns} & - & & & & & \\
        \footnotesize\textbf{\texttt{Opt-Sub-Rat}} & \footnotesize{ns} & \footnotesize{ns} & \footnotesize{ns} & - & & & & \\
        \footnotesize\textbf{\texttt{Pes-Aut-Emo}} & \cellcolor{gray!10}\footnotesize{<.01} & \cellcolor{gray!10}\footnotesize{<.001} & \cellcolor{gray!10}\footnotesize{<.01} & \cellcolor{gray!10}\footnotesize{<.0001} & - & & & \\
        \footnotesize\textbf{\texttt{Pes-Aut-Rat}} & \footnotesize{ns} & \footnotesize{ns} & \footnotesize{ns} & \cellcolor{gray!10}\footnotesize{.03} & \footnotesize{ns} & - & & \\
        \footnotesize\textbf{\texttt{Pes-Sub-Emo}} & \cellcolor{gray!10}\footnotesize{.03} & \cellcolor{gray!10}\footnotesize{.01} & \cellcolor{gray!10}\footnotesize{.04} & \cellcolor{gray!10}\footnotesize{<.001} & \footnotesize{ns} & \footnotesize{ns} & - & \\
        \footnotesize\textbf{\texttt{Pes-Sub-Rat}} & \cellcolor{gray!10}\footnotesize{<.0001} & \cellcolor{gray!10}\footnotesize{<.0001} & \cellcolor{gray!10}\footnotesize{<.001} & \cellcolor{gray!10}\footnotesize{<.0001} & \footnotesize{ns} & \footnotesize{ns} & \footnotesize{ns} & - \\
    \bottomrule
        % F(5, 338) = 2.32, p = .04, $R^2$ = 0.019 & & & & \\
    % \bottomrule
    \end{tabular}}}
    \label{tab:fig_4a}
\end{table*}

%% Fig 4b (Valence Towards Cause ~ CA Condition)
\begin{table*}[!hbt]
    \caption{Pairwise comparisons of participants' perceived \textbf{valence (towards the cause)} between the different \textbf{CA conditions} (see Figure~\ref{fig:other-valence}). The adjusted \textit{p}-values are computed using Bonferroni correction.}
    \centering
    {\new{\begin{tabular}{r c c c c c c c c}
    \toprule
        & \footnotesize\textbf{\texttt{Opt-Aut-Emo}} & \footnotesize\textbf{\texttt{Opt-Aut-Rat}} & \footnotesize\textbf{\texttt{Opt-Sub-Emo}} & \footnotesize\textbf{\texttt{Opt-Sub-Rat}} & \footnotesize\textbf{\texttt{Pes-Aut-Emo}} & \footnotesize\textbf{\texttt{Pes-Aut-Rat}} & \footnotesize\textbf{\texttt{Pes-Sub-Emo}} & \footnotesize\textbf{\texttt{Pes-Sub-Rat}} \\
    \midrule
        \footnotesize\textbf{\texttt{Opt-Aut-Emo}} & - & & & & & & & \\
        \footnotesize\textbf{\texttt{Opt-Aut-Rat}} & \footnotesize{ns} & - & & & & & & \\
        \footnotesize\textbf{\texttt{Opt-Sub-Emo}} & \footnotesize{ns} & \footnotesize{ns} & - & & & & & \\
        \footnotesize\textbf{\texttt{Opt-Sub-Rat}} & \footnotesize{ns} & \footnotesize{ns} & \footnotesize{ns} & - & & & & \\
        \footnotesize\textbf{\texttt{Pes-Aut-Emo}} & \cellcolor{gray!10}\footnotesize{<.0001} & \cellcolor{gray!10}\footnotesize{.02} & \cellcolor{gray!10}\footnotesize{.001} & \cellcolor{gray!10}\footnotesize{<.001} & - & & & \\
        \footnotesize\textbf{\texttt{Pes-Aut-Rat}} & \cellcolor{gray!10}\footnotesize{.04} & \footnotesize{ns} & \footnotesize{ns} & \footnotesize{ns} & \footnotesize{ns} & - & & \\
        \footnotesize\textbf{\texttt{Pes-Sub-Emo}} & \cellcolor{gray!10}\footnotesize{<.01} & \footnotesize{ns} & \cellcolor{gray!10}\footnotesize{.05} & \cellcolor{gray!10}\footnotesize{.03} & \footnotesize{ns} & \footnotesize{ns} & - & \\
        \footnotesize\textbf{\texttt{Pes-Sub-Rat}} & \cellcolor{gray!10}\footnotesize{<.0001} & \cellcolor{gray!10}\footnotesize{<.001} & \cellcolor{gray!10}\footnotesize{<.0001} & \cellcolor{gray!10}\footnotesize{<.0001} & \footnotesize{ns} & \cellcolor{gray!10}\footnotesize{.03} & \footnotesize{ns} & - \\
    \bottomrule
        % F(5, 338) = 2.32, p = .04, $R^2$ = 0.019 & & & & \\
    % \bottomrule
    \end{tabular}}}
    \label{tab:fig_4b}
\end{table*}

%% Fig 4c (Arousal Towards Self ~ CA Condition)
\begin{table*}[!hbt]
    \caption{Pairwise comparisons of participants' perceived \textbf{arousal (towards self)} between the different \textbf{CA conditions} (see Figure~\ref{fig:self-arousal}). The adjusted \textit{p}-values are computed using Bonferroni correction.}
    \centering
    {\new{\begin{tabular}{r c c c c c c c c}
    \toprule
        & \footnotesize\textbf{\texttt{Opt-Aut-Emo}} & \footnotesize\textbf{\texttt{Opt-Aut-Rat}} & \footnotesize\textbf{\texttt{Opt-Sub-Emo}} & \footnotesize\textbf{\texttt{Opt-Sub-Rat}} & \footnotesize\textbf{\texttt{Pes-Aut-Emo}} & \footnotesize\textbf{\texttt{Pes-Aut-Rat}} & \footnotesize\textbf{\texttt{Pes-Sub-Emo}} & \footnotesize\textbf{\texttt{Pes-Sub-Rat}} \\
    \midrule
        \footnotesize\textbf{\texttt{Opt-Aut-Emo}} & - & & & & & & & \\
        \footnotesize\textbf{\texttt{Opt-Aut-Rat}} & \footnotesize{ns} & - & & & & & & \\
        \footnotesize\textbf{\texttt{Opt-Sub-Emo}} & \footnotesize{ns} & \footnotesize{ns} & - & & & & & \\
        \footnotesize\textbf{\texttt{Opt-Sub-Rat}} & \footnotesize{ns} & \footnotesize{ns} & \footnotesize{ns} & - & & & & \\
        \footnotesize\textbf{\texttt{Pes-Aut-Emo}} & \footnotesize{ns} & \footnotesize{ns} & \footnotesize{ns} & \footnotesize{ns} & - & & & \\
        \footnotesize\textbf{\texttt{Pes-Aut-Rat}} & \footnotesize{ns} & \footnotesize{ns} & \footnotesize{ns} & \footnotesize{ns} & \footnotesize{ns} & - & & \\
        \footnotesize\textbf{\texttt{Pes-Sub-Emo}} & \footnotesize{ns} & \footnotesize{ns} & \footnotesize{ns} & \footnotesize{ns} & \footnotesize{ns} & \footnotesize{ns} & - & \\
        \footnotesize\textbf{\texttt{Pes-Sub-Rat}} & \footnotesize{ns} & \footnotesize{ns} & \footnotesize{ns} & \footnotesize{ns} & \footnotesize{ns} & \footnotesize{ns} & \footnotesize{ns} & - \\
    \bottomrule
        % F(5, 338) = 2.32, p = .04, $R^2$ = 0.019 & & & & \\
    % \bottomrule
    \end{tabular}}}
    \label{tab:fig_4c}
\end{table*}

%% Fig 4d (Arousal Towards Cause ~ CA Condition)
\begin{table*}[!hbt]
    \caption{Pairwise comparisons of participants' perceived \textbf{arousal (towards the cause)} between the different \textbf{CA conditions} (see Figure~\ref{fig:other-arousal}). The adjusted \textit{p}-values are computed using Bonferroni correction.}
    \centering
    {\new{\begin{tabular}{r c c c c c c c c}
    \toprule
        & \footnotesize\textbf{\texttt{Opt-Aut-Emo}} & \footnotesize\textbf{\texttt{Opt-Aut-Rat}} & \footnotesize\textbf{\texttt{Opt-Sub-Emo}} & \footnotesize\textbf{\texttt{Opt-Sub-Rat}} & \footnotesize\textbf{\texttt{Pes-Aut-Emo}} & \footnotesize\textbf{\texttt{Pes-Aut-Rat}} & \footnotesize\textbf{\texttt{Pes-Sub-Emo}} & \footnotesize\textbf{\texttt{Pes-Sub-Rat}} \\
    \midrule
        \footnotesize\textbf{\texttt{Opt-Aut-Emo}} & - & & & & & & & \\
        \footnotesize\textbf{\texttt{Opt-Aut-Rat}} & \footnotesize{ns} & - & & & & & & \\
        \footnotesize\textbf{\texttt{Opt-Sub-Emo}} & \footnotesize{ns} & \footnotesize{ns} & - & & & & & \\
        \footnotesize\textbf{\texttt{Opt-Sub-Rat}} & \footnotesize{ns} & \footnotesize{ns} & \footnotesize{ns} & - & & & & \\
        \footnotesize\textbf{\texttt{Pes-Aut-Emo}} & \cellcolor{gray!10}\footnotesize{<.01} & \cellcolor{gray!10}\footnotesize{.01} & \footnotesize{ns} & \footnotesize{ns} & - & & & \\
        \footnotesize\textbf{\texttt{Pes-Aut-Rat}} & \footnotesize{ns} & \footnotesize{ns} & \footnotesize{ns} & \footnotesize{ns} & \footnotesize{ns} & - & & \\
        \footnotesize\textbf{\texttt{Pes-Sub-Emo}} & \footnotesize{ns} & \footnotesize{ns} & \footnotesize{ns} & \footnotesize{ns} & \footnotesize{ns} & \footnotesize{ns} & - & \\
        \footnotesize\textbf{\texttt{Pes-Sub-Rat}} & \cellcolor{gray!10}\footnotesize{<.0001} & \cellcolor{gray!10}\footnotesize{<.001} & \cellcolor{gray!10}\footnotesize{.03} & \cellcolor{gray!10}\footnotesize{.02} & \footnotesize{ns} & \footnotesize{ns} & \footnotesize{ns} & - \\
    \bottomrule
        % F(5, 338) = 2.32, p = .04, $R^2$ = 0.019 & & & & \\
    % \bottomrule
    \end{tabular}}}
    \label{tab:fig_4d}
\end{table*}

%% Fig 4e (Dominance Towards Self ~ CA Condition
\begin{table*}[!hbt]
    \caption{Pairwise comparisons of participants' perceived \textbf{dominance (towards self)} between the different \textbf{CA conditions} (see Figure~\ref{fig:self-dominance}). The adjusted \textit{p}-values are computed using Bonferroni correction.}
    \centering
    {\new{\begin{tabular}{r c c c c c c c c}
    \toprule
        & \footnotesize\textbf{\texttt{Opt-Aut-Emo}} & \footnotesize\textbf{\texttt{Opt-Aut-Rat}} & \footnotesize\textbf{\texttt{Opt-Sub-Emo}} & \footnotesize\textbf{\texttt{Opt-Sub-Rat}} & \footnotesize\textbf{\texttt{Pes-Aut-Emo}} & \footnotesize\textbf{\texttt{Pes-Aut-Rat}} & \footnotesize\textbf{\texttt{Pes-Sub-Emo}} & \footnotesize\textbf{\texttt{Pes-Sub-Rat}} \\
    \midrule
        \footnotesize\textbf{\texttt{Opt-Aut-Emo}} & - & & & & & & & \\
        \footnotesize\textbf{\texttt{Opt-Aut-Rat}} & \footnotesize{ns} & - & & & & & & \\
        \footnotesize\textbf{\texttt{Opt-Sub-Emo}} & \footnotesize{ns} & \footnotesize{ns} & - & & & & & \\
        \footnotesize\textbf{\texttt{Opt-Sub-Rat}} & \footnotesize{ns} & \footnotesize{ns} & \footnotesize{ns} & - & & & & \\
        \footnotesize\textbf{\texttt{Pes-Aut-Emo}} & \footnotesize{ns} & \footnotesize{ns} & \footnotesize{ns} & \footnotesize{ns} & - & & & \\
        \footnotesize\textbf{\texttt{Pes-Aut-Rat}} & \footnotesize{ns} & \footnotesize{ns} & \footnotesize{ns} & \footnotesize{ns} & \footnotesize{ns} & - & & \\
        \footnotesize\textbf{\texttt{Pes-Sub-Emo}} & \footnotesize{ns} & \footnotesize{ns} & \footnotesize{ns} & \footnotesize{ns} & \footnotesize{ns} & \footnotesize{ns} & - & \\
        \footnotesize\textbf{\texttt{Pes-Sub-Rat}} & \footnotesize{ns} & \cellcolor{gray!10}\footnotesize{.03} & \footnotesize{ns} & \cellcolor{gray!10}\footnotesize{.02} & \footnotesize{ns} & \footnotesize{ns} & \footnotesize{ns} & - \\
    \bottomrule
        % F(5, 338) = 2.32, p = .04, $R^2$ = 0.019 & & & & \\
    % \bottomrule
    \end{tabular}}}
    \label{tab:fig_4e}
\end{table*}

%% Fig 4f (Dominance Towards Cause ~ CA COndition)
\begin{table*}[!hbt]
    \caption{Pairwise comparisons of participants' perceived \textbf{dominance (towards the cause)} between the different \textbf{CA conditions} (see Figure~\ref{fig:other-dominance}). The adjusted \textit{p}-values are computed using Bonferroni correction.}
    \centering
    {\new{\begin{tabular}{r c c c c c c c c}
    \toprule
        & \footnotesize\textbf{\texttt{Opt-Aut-Emo}} & \footnotesize\textbf{\texttt{Opt-Aut-Rat}} & \footnotesize\textbf{\texttt{Opt-Sub-Emo}} & \footnotesize\textbf{\texttt{Opt-Sub-Rat}} & \footnotesize\textbf{\texttt{Pes-Aut-Emo}} & \footnotesize\textbf{\texttt{Pes-Aut-Rat}} & \footnotesize\textbf{\texttt{Pes-Sub-Emo}} & \footnotesize\textbf{\texttt{Pes-Sub-Rat}} \\
    \midrule
        \footnotesize\textbf{\texttt{Opt-Aut-Emo}} & - & & & & & & & \\
        \footnotesize\textbf{\texttt{Opt-Aut-Rat}} & \footnotesize{ns} & - & & & & & & \\
        \footnotesize\textbf{\texttt{Opt-Sub-Emo}} & \footnotesize{ns} & \footnotesize{ns} & - & & & & & \\
        \footnotesize\textbf{\texttt{Opt-Sub-Rat}} & \footnotesize{ns} & \footnotesize{ns} & \footnotesize{ns} & - & & & & \\
        \footnotesize\textbf{\texttt{Pes-Aut-Emo}} & \footnotesize{ns} & \footnotesize{ns} & \footnotesize{ns} & \footnotesize{ns} & - & & & \\
        \footnotesize\textbf{\texttt{Pes-Aut-Rat}} & \footnotesize{ns} & \footnotesize{ns} & \footnotesize{ns} & \footnotesize{ns} & \footnotesize{ns} & - & & \\
        \footnotesize\textbf{\texttt{Pes-Sub-Emo}} & \footnotesize{ns} & \footnotesize{ns} & \footnotesize{ns} & \footnotesize{ns} & \footnotesize{ns} & \footnotesize{ns} & - & \\
        \footnotesize\textbf{\texttt{Pes-Sub-Rat}} & \cellcolor{gray!10}\footnotesize{<.01} & \footnotesize{ns} & \footnotesize{ns} & \footnotesize{ns} & \footnotesize{ns} & \footnotesize{ns} & \footnotesize{ns} & - \\
    \bottomrule
        % F(5, 338) = 2.32, p = .04, $R^2$ = 0.019 & & & & \\
    % \bottomrule
    \end{tabular}}}
    \label{tab:fig_4f}
\end{table*}
% \newpage

%TC:endignore

\end{document}